\documentclass[11pt,a4paper]{article}                           

\usepackage[bookmarksopen, bookmarksnumbered, bookmarksopenlevel=2]{hyperref}
\usepackage{tikz}
\usepackage{tikz-3dplot}
\usetikzlibrary{calc}
\usetikzlibrary{decorations} %
\usepackage[UKenglish]{babel}
\usepackage[toc,page]{appendix}
\usepackage{amsmath}
\usepackage{amssymb}
\usepackage{graphicx}
\usepackage{hhline}
\usepackage[bf]{caption}
\usepackage{cite}
\usepackage[vcentermath]{youngtab}
\usepackage{geometry}
\DeclareSymbolFont{bbold}{U}{bbold}{m}{n}
\DeclareSymbolFontAlphabet{\mathbbold}{bbold}

\newcommand{\un}{\mathbf{1}}

\newcommand{\f}{\mathbf{5}}
\newcommand{\fb}{\mathbf{\bar{5}}}
\newcommand{\te}{\mathbf{10}}
\newcommand{\teb}{\mathbf{\bar{10}}}

\usepackage{cancel}

\hypersetup{
    pdftitle={Discrete Gauge Symmetries by Higgsing in Four-Dimensional F-Theory Compactifications},
    pdfauthor={Christoph Mayrhofer, Eran Palti, Oskar Till, Timo Weigand},
    pdfsubject={F-theory compactifications}
}
\numberwithin{equation}{section}
\numberwithin{table}{section}

\newcommand{\bea}{\begin{eqnarray}}
\newcommand{\eea}{\end{eqnarray}}

\newcommand{\tw}{{\rm w}}

\newcommand{\executeiffilenewer}[3]{%
 \ifnum\pdfstrcmp{\pdffilemoddate{#1}}%
 {\pdffilemoddate{#2}}>0%
 {\immediate\write18{#3}}\fi%
}

\begin{document}

\baselineskip=14pt
\parskip 5pt plus 1pt

\vspace*{-1.5cm}
\begin{flushright}    
  {\small

  }
\end{flushright}

\vspace{2cm}
\begin{center}        
  {\LARGE Discrete Gauge Symmetries by Higgsing in Four-Dimensional F-Theory Compactifications}
\end{center}

\vspace{0.75cm}
\begin{center}        
Christoph Mayrhofer, Eran Palti, Oskar Till and Timo Weigand
\end{center}

\vspace{0.15cm}
\begin{center}        
\emph{Institut f\"ur Theoretische Physik, Ruprecht-Karls-Universit\"at, \\
             Philosophenweg 19, 69120,
             Heidelberg, Germany\\
             }
\end{center}

\vspace{2cm}


\begin{abstract}
\noindent We study F-theory compactifications to four dimensions that exhibit discrete gauge symmetries. Geometrically these arise by deforming elliptic fibrations with two sections to a genus-one fibration with a bi-section. From a four-dimensional field theory perspective they are remnant symmetries from a Higgsed $U(1)$ gauge symmetry. We implement such symmetries in the presence of an additional $SU(5)$ symmetry and associated matter fields, giving a geometric prescription for calculating the induced discrete charge for the matter curves and showing the absence of Yukawa couplings that are forbidden by this charge. We present a detailed map between the field theory and the geometry, including an identification of the Higgs field and the massless states before and after the Higgsing. Finally we show that the Higgsing of the $U(1)$ induces a G-flux which precisely accounts for the change in the Calabi-Yau Euler number so as to leave the D3 tadpole invariant.
\end{abstract}

\thispagestyle{empty}
\clearpage
\tableofcontents
\thispagestyle{empty}


\newpage
\setcounter{page}{1}

\section{Introduction}

F-theory provides a reformulation of particle physics in terms of the geometry of varieties. In particular four-dimensional supersymmetric particle physics is mapped to the geometry of Calabi-Yau fourfolds. The geometric construction of non-abelian gauge symmetries and charged matter has been well understood since the original work \cite{Vafa:1996xn,Morrison:1996na,Morrison:1996pp}, while more recently abelian gauge symmetries have been under intensive investigation in both four and six dimensions, see \cite{Grimm:2010ez,Grimm:2011tb,Braun:2011zm,Krause:2011xj,Grimm:2011fx,Krause:2012yh,Morrison:2012ei,Cvetic:2012xn,   Mayrhofer:2012zy,Braun:2013yti,Borchmann:2013jwa,Cvetic:2013nia,Braun:2013nqa,Cvetic:2013uta,Borchmann:2013hta,Cvetic:2013jta,Cvetic:2013qsa,Mayrhofer:2013ara,Krippendorf:2014xba,Braun:2014nva,Morrison:2014era,Martini:2014iza,Bizet:2014uua,Kuntzler:2014ila,Klevers:2014bqa}  for some recent work and \cite{Denef:2008wq,Weigand:2010wm,Maharana:2012tu} for reviews. In this article we will study the manifestation of discrete gauge symmetries in four-dimensional models. Discrete gauge symmetries are remnants of broken continuous gauge symmetries. They share the property with their continuous origin of constraining the operator spectrum of the theory, but do not have a massless propagating degree of freedom associated to them. They have been widely studied from a field theory perspective both formally and with phenomenological applications, and have also been actively studied within a string theory context, for recent work see  \cite{BerasaluceGonzalez:2011wy,Ibanez:2012wg,BerasaluceGonzalez:2012vb,BerasaluceGonzalez:2012zn,Marchesano:2013ega,Honecker:2013hda,Berasaluce-Gonzalez:2013bba,Antoniadis:2013joa,Karozas:2014aha}. Perhaps the most direct way to construct discrete gauge symmetries is to start with a theory with a continuous gauge symmetry and Higgs it with a field that is neutral under some discrete subgroup of the gauge group. In this paper we will study a realisation of this mechanism in F-theory.\footnote{For recent work studying the Higgsing of abelian symmetries in smooth heterotic string models see \cite{Buchbinder:2013dna,Buchbinder:2014qda,Buchbinder:2014sya}. The Higgsing of non-abelian gauge symmetries in F-theory compactifications via deformations has been described in detail in \cite{Grassi:2013kha,Grassi:2014sda}.}

Our starting point is the fibration considered in \cite{Morrison:2012ei}, which is the most general form of the elliptic fibration that supports two sections, a so-called zero section and an additional section which can be associated to a massless $U(1)$ gauge symmetry. The fibre is given in terms of a (restricted) quartic in the projective space $\textmd{Bl}^1\mathbb{P}_{1,1,2}$,
\begin{equation}
\begin{aligned} \label{eq:Bl1124intro}
 s w^2   + b_{0}  s^2 u^2 w  + b_1s u v w  + b_2 v^2 w + c_{0}s^3 u^4   + c_{1} s^2  u^3 v  + c_{2}s u^2 v^2   + c_3 u v^3 = 0\;.
\end{aligned}
\end{equation}
Here $b_i$ and $c_i$ are sections of suitable line bundles on the base ${\cal B}$ of the fibration, $u$, $v$ and $w$ are the projective coordinates on $\mathbb{P}_{1,1,2}$, and $s$ is the projective coordinate related to the exceptional divisor of the blow-up of $w=u=0$. As was shown in \cite{Morrison:2012ei} this fibration possesses $I_2$ fibres over codimension-two loci in the base where matter charged under the $U(1)$ symmetry resides. In particular there are two types of matter fields, $\mathbf{1}_{1}$ and $\mathbf{1}_{2}$, with the subscript denoting their charge, which reside over two distinct loci in the base. It was shown in \cite{Braun:2014oya} that deforming this equation, by first blowing down the divisor corresponding to $s$ and then turning on an additional monomial $c_4 v^4$, combines the two sections into a single bi-section. The resulting geometry no longer supports a zero section. The implication of this is that the associated Jacobian fibration, which is the Weierstrass model associated to $\mathbb{P}_{1,1,2}[4]$ through a birational coordinate transformation, is singular and does not admit a Calabi-Yau resolution \cite{Braun:2014oya}. However the $\mathbb{P}_{1,1,2}[4]$ fibration is perfectly smooth and allows for a controlled analysis of the geometry and the resulting physics. In particular, in \cite{Morrison:2014era} the interpretation of this deformation was studied further and described in terms of the Higgsing of the field with charge 2 under the $U(1)$ symmetry, $\mathbf{1}_{2}$, which breaks it to a $\mathbb{Z}_2$ discrete group. The fibration was studied in \cite{Morrison:2014era}  on explicit complex two-dimensional bases and the resulting six-dimensional theories were described. A slightly more general version of the $\mathbb{Z}_2$ fibration of \cite{Braun:2014oya,Morrison:2014era} over complex two-dimensional bases   was studied  subsequently in \cite{Anderson:2014yva}. 
Note that the appearance of a discrete gauge group factor as a result of a multi-section in the F-theory fibration is not to be confused with the quotienting of the gauge group by a discrete subgroup as is the consequence of a torsional element of the Mordell-Weil group \cite{Aspinwall:1998xj,Mayrhofer:2014opa}.

In this paper we will study implications of the  $\mathbb{Z}_2$ discrete group for four-dimensional F-theory compactifications corresponding to base spaces of complex dimension three.
Amongst other things we will, in addition to the discrete group $\mathbb{Z}_2$, implement a further non-abelian gauge symmetry. As an example we will choose several embeddings of $SU(5)$. There are a number of qualitatively new effects when going from six to four dimensions. 
The first is that in the four-dimensional models Yukawa couplings between the charged fields arise which are localised at points of codimension three in the base.
 In the presence of the $SU(5)$ singularity we have five types of such couplings with respect to their $SU(5)$ representations:  $\fb \,  \fb \, \te$, $\f  \,  \te \,  \te $,  $\un \, \fb \, \f $, $\un \, \te \,  {\teb} $, $\un \,  \un \, \un$. The fibration \eqref{eq:Bl1124intro}  with the massless $U(1)$ symmetry has already been studied in the presence of an additional $SU(5)$ singularity in \cite{Borchmann:2013hta}, where it was shown that all such couplings which conserve the $U(1)$ charge appear in the geometry. From a four-dimensional field theory perspective, Higgsing such a theory with the field $\mathbf{1}_{2}$ is expected to lead to two primary, related,  effects. The first is that one expects the selection rules governing the presence of a coupling to be modified from the $U(1)$ charge to a $\mathbb{Z}_2$ charge. The second is that some fields can gain a mass from operators of the type $\un \, \fb \, \f $, $\un\, \te  \, \teb $, $\un\,   \un \, \un$ which involve the Higgsed $SU(5)$ singlet field. The main aim of the paper is to study the geometric manifestation of these effects. We will show that the deformation of the geometry recombines the matter curves which participate in couplings with the Higgsed singlet. Thereby it will change the massless spectrum on them. In particular this recombination will be such that one can associate a $\mathbb{Z}_2$ charge to the recombined matter curves and that the presence of a coupling between the fields is dictated by their $\mathbb{Z}_2$ charge.

We will provide further evidence for this picture by analysing the four-dimensional gauge theory compactified on the M-theory circle. 
A key observation is that the Higgs field responsible for the deformation is to be identified with a specific Kaluza-Klein (KK) mode of the states ${\mathbf{1}_2}$ with respect to this circle reduction. 
By identifying the correct KK mode in the fibre geometry we are able to further support and, in fact, extend the field picture suggested in \cite{Morrison:2014era}. This understanding also explains some of the geometric data of the conifold transition studied for concrete six-dimensional examples in \cite{Anderson:2014yva}.

Another qualitatively new effect in four-dimensional models is the presence of $G_4$-flux in the internal manifold. This plays an essential role in the Higgsing process. Since the Higgsing gives a mass to a $U(1)$ gauge field it must reflect in a change in the cohomology of the Calabi-Yau fourfold. Indeed the singular locus associated to the charged singlet $\mathbf{1}_{2}$ is a curve of conifold singularities, and the geometric transition corresponding to their Higgsing, of blowing down and then deforming, is just a conifold transition over this curve. A particular consequence of this is that the Euler number of the manifold changes. However the Euler number is known to appear in the D3 tadpole condition
\begin{eqnarray}
\int_{CY_4} G_4 \wedge G_4 = \frac{\chi\left( CY_4 \right)}{12} \;.
\end{eqnarray}
Therefore, in order to maintain a tadpole-free configuration, new $G_4$-flux must be induced by the transition \cite{Braun:2011zm,Krause:2012yh,Intriligator:2012ue}. We will present this flux and show that it can account for the associated change in the Euler number.

An important motivation for this work is to develop the technical tools and an understanding which will allow us to implement discrete gauge symmetries obtained by Higgsing in F-theory model building. Indeed, the particular geometries studied in this paper, being Grand Unified Theories (GUTs) with a remnant discrete gauge symmetry, are also of interest from a phenomenological perspective. With this in mind we will present two examples where for one of the two models the remnant  $\mathbb{Z}_2$ symmetry can be identified with the R-parity of the Minimal Supersymmetric Standard Model (MSSM).

The layout of the paper is as follows. In section~\ref{sec:U(1)fibration}  and \ref{sec:Z2fibration} we describe the elliptic and torus fibrations with gauge group $U(1)$ and $\mathbb Z_2$, respectively, with special emphasis on fibrations over bases of complex dimension three. We work out the singlet curves with the help of a prime ideal decomposition and analyse their intersection loci in codimension three. In section \ref{sec:compensator_G4-flux} we provide a candidate for a $G_4$-flux for the bi-section model which compensates the change in the D3-brane number induced by the deformation from the $U(1)$ to the $\mathbb Z_2$-model. 
In section~\ref{sec:the_field_theory_picture} we interpret the conifold transition as a Higgsing due to a specific KK state in the circle reduction from four to three dimensions and match this field theoretic picture to the geometry. 
In section~\ref{sec:su5xu1-versus-su5Z2} we introduce an additional $SU(5)$ gauge group, resulting in F-theory compactifications with gauge group $SU(5) \times U(1)$ and $SU(5) \times \mathbb Z_2$, respectively. In particular we demonstrate the nature of the conifold transition as a Higgsing by inspecting the change of the charged matter curves and the structure of selection rules in the Yukawa sector. 
As a further application, we present in section \ref{sec:R-parity} another $SU(5) \times \mathbb Z_2$ GUT  model in which the $\mathbb Z_2$ group corresponds to R-parity. A detailed derivation of the remnant discrete gauge group factor after Higgsing a $U(1)$ in the presence of extra non-abelian gauge symmetries is presented in appendix \ref{App-disc}. We end this article with our conclusions in section~\ref{sec:conclusions}.

\noindent{\bf Note added}: As this work was being completed, the two papers \cite{Klevers:2014bqa,Garcia-Etxebarria:2014qua} appeared, which have partial overlap with our results.

\section{Fibrations with gauge group \texorpdfstring{$U(1)$ and $\mathbb Z_2$}{U(1) and Z2}}

\subsection{\texorpdfstring{${\rm Bl}^1 \mathbb P_{1,1,2}[4]$}{Bl1P112[4]}-fibrations over 3-dimensional bases with \texorpdfstring{$U(1)$}{U(1)} gauge group} \label{sec:U(1)fibration}

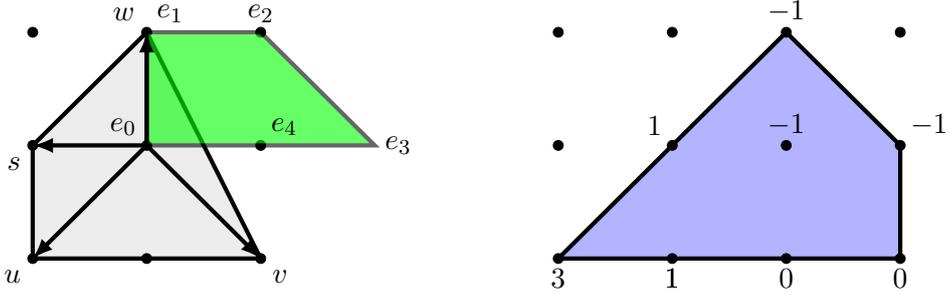
\begin{figure}[t]
    \centering
  \begin{tikzpicture}[scale=1.5]
  \filldraw [ultra thick, draw=black, fill=lightgray!30!white]
      (-1,-1)--(1,-1)--(0,1)--(-1,0)--cycle;
    \foreach \x in {-1,0,1}{
      \foreach \y in {-1,0,...,1}{
        \node[draw,circle,inner sep=1.3pt,fill] at (\x,\y) {};
      }
    }
              \filldraw [ultra thick, draw=black, fill=green, opacity=0.6]
      (0,0)--(2,0)--(1,1)--(0,1)--cycle;
  \node [above left] at (0,0) {$e_0$};
    \node [above right] at (0,1) {$e_1$};
  \node [above] at (1,1) {$e_2$};
  \node [right] at (2,0) {$e_3$};
  \node [above right] at (1,0) {$e_4$};

  \draw[ultra thick, -latex]
       (0,0) -- (0,1) node[above left] {$w$};
  \draw[ultra thick, -latex]
       (0,0) -- (1,-1) node[below right] {$v$};
  \draw[ultra thick, -latex]
       (0,0) -- (-1,-1) node[below left] {$u$};
         \draw[ultra thick, -latex]
       (0,0) -- (-1,0) node[below left] {$s$};
\begin{scope}[xshift=0.33\textwidth]
\filldraw [ultra thick, draw=black, fill=blue!30!white]
      (0,1)--(-2,-1)--(1,-1)--(1,0)--cycle;
    \foreach \x in {-2,-1,...,1}{
      \foreach \y in {-1,0,...,1}{
        \node[draw,circle,inner sep=1.3pt,fill] at (\x,\y) {};
      }
    }
  \node [above] at (0,0)  {$-1$};
  \node [above] at (0,1) {$-1$};
  \node [above right] at (1,0) {$-1$};
      \node [below] at (1,-1) {$0$};
        \node [below] at (0,-1) {$0$};
          \node [below] at (-1,-1) {$1$};
   \node [below] at (-1,-1) {$1$};
   \node [below] at (-2,-1) {$3$};
  \node [above left] at(-1,0)  {$1$};
\end{scope}
  \end{tikzpicture}
      \caption{$SU(5)$ top $2$ over polygon 6 of \cite{Bouchard:2003bu} together with its dual polygon, bounded below by the values $z_{min}$, shown next to the nodes.}\label{fig:polygon6}
\end{figure}

The starting point of our analysis is what was derived in \cite{Morrison:2012ei} as the most general elliptic fibration with Mordell-Weil group of rank $1$ over a \emph{generic} base space ${\cal B}$. Such geometries are described as a ${\rm Bl}^1 \mathbb P_{1,1,2}[4]$-fibration over ${\cal B}$, corresponding to the vanishing locus of the hypersurface equation
\begin{equation}
\begin{aligned} \label{eq:Bl1124}
P_1: = s w^2   + b_{0}  s^2 u^2 w  + b_1s u v w  + b_2 v^2 w 
 + c_{0}s^3 u^4   + c_{1} s^2  u^3 v  + c_{2}s u^2 v^2   + c_3 u v^3
\end{aligned}
\end{equation}
for $b_i$ and $c_i$ sections of suitable line bundles on ${\cal B}$.
The coordinates $[u : v : w: s]$ span the toric ambient space of the elliptic fibre with scaling relations 
\begin{eqnarray}
(u,v,w,s) \simeq (\lambda u, \lambda \mu v, \lambda^2  \mu \, w, \mu \, s)
\end{eqnarray}
and Stanley-Reisner ideal $\{  u \, w, v \, s\}$.
This fibre corresponds to what was called polygon 6 in \cite{Bouchard:2003bu}, where all toric tops over the 16 hypersurface representations of genus-one curves in toric ambient spaces have been classified. We will adopt their enumeration, since we are applying their top construction in section \ref{sec:su5xu1-versus-su5Z2} and \ref{sec:R-parity}. To describe a Calabi-Yau fibration, one has the freedom to introduce a line bundle ${\cal L}$ on ${\cal B}$ with first Chern class $c_1({\cal L}) =: \beta$. 
Then, the fibre coordinates transform as sections of bundles 
as  summarised in Table~\eqref{tab:scaling}, while the coefficients $b_i$ and $c_i$ transform as sections of line bundles on ${\cal B}$ whose first Chern classes are displayed in Table~\ref{tab:coeff1}.

\begin{table}
\centering
\begin{tabular}{c||cccccc}
 & $u$ & $v $& $w$ & $s$\\
\hline
\hline
$\alpha = \beta - \bar {\cal K}$ & $\cdot$ & 1 & $\cdot$ & $\cdot$  \\
$\beta$ & $\cdot$ & $\cdot$ & 1 &  $\cdot$ \\
\hline
U & 1 & 1 & 2 & $\cdot$ \\
S & $\cdot$ & 1 & 1 & 1 
\end{tabular}
\caption{Divisor classes and coordinates of the fibre ambient space, with ${\cal K}$ the canonical bundle of the base ${\cal B}$ and $\beta$ an 'arbitrary' class on ${\cal B}$.}\label{tab:scaling}
\end{table}

\begin{table} 
\centering
\begin{tabular}{c|c|c|c|c|c|c|c}
 $b_{0}$ & $b_{1}$ & $b_{2}$ & $c_{0}$ & $c_1$ & $c_2$&$c_{3}$ & $c_4$ \\
\hline
 $\beta$& $\bar {\cal K}$ &  $-\beta+2 \bar {\cal K}$ &   $ 2 \beta$ &   $\beta + \bar {\cal K}$ &    $ 2\bar {\cal K}$ &   $-\beta+3\bar {\cal K}$ &   $ -2\beta+4 \bar {\cal K}$\\
\end{tabular}
\caption{Classes of the coefficients entering \eqref{eq:Bl1124}.}\label{tab:coeff1}
\end{table}

If we take \eqref{eq:Bl1124} and set $s \equiv 1$, we obtain the polynomial of a non-generic quartic in $\mathbb P_{1,1,2}$. 
This fibration exhibits a curve of conifold singularities situated at $u = w=0$ in the fibre over the curve $b_2 = c_3=0$. The appearance of this curve of singularities is a consequence of the non-generic form of the hypersurface as a quartic in $\mathbb P_{1,1,2}$ with coefficient of the term $v^4$ set to zero. The absence of this term is responsible for the appearance of two independent sections in the singular model which generate a Mordell-Weil group of rank one,
\begin{eqnarray} \label{eq:sec01}
 {\rm Sec}_1: [u : v : w] = [0 : 1 : -b_2], \qquad {\rm Sec}_2: [u : v : w] = [0 : 1 : 0].
\end{eqnarray}
The section ${\rm Sec}_1$ intersects the conifold points over the curve $b_2 = c_3=0$. 
After the blow-up resolution of this conifold point, $u \rightarrow u \, s, w \rightarrow w \, s$, ${\rm Sec}_1$ is replaced by the resolution divisor  $S: s=0$, which forms a rational section.
The holomorphic zero-section of the fibration \eqref{eq:Bl1124}  is identified with the divisor $U: u=0$. 
The image of $S$ under the Shioda map, 
\begin{eqnarray} \label{eq:Shiodamap}
\tw = S-U-\bar{\mathcal{K}}-[b_2],
\end{eqnarray}
acts as the generator of a $U(1)$ gauge symmetry in F-theory compactifications on the elliptic fibration. 

This geometry has been analysed in great detail in \cite{Morrison:2012ei} for a generic base space ${\cal B}$  of complex dimension two. 
Since the essential features of Higgsing and the resulting discrete selection rules are visible only in compactifications of F-theory to four dimensions, we must first extend this analysis to base spaces ${\cal B}$ of complex dimension three. The novel feature compared to \cite{Morrison:2012ei} is the appearance of 
a more complicated structure of the singlet loci as in general singular \emph{curves} (as opposed to points) on ${\cal B}$ and their intersection at Yukawa points in codimension three on ${\cal B}$.

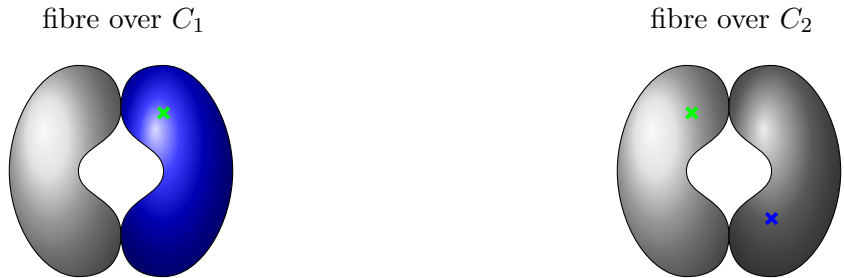
\begin{figure}
\centering
\begin{tikzpicture}

\node at (1.3,2) {fibre over $C_1$};

\begin{scope}[scale=0.7]
   \shadedraw[ball color=gray!30!white] (-0.3,0) .. controls (-0.3,1) and (.3,2) .. (1,2)
               .. controls (1.7,2) and (1.8,1.5) .. (1.8,1.2)
               .. controls (1.8,.5) and (1,.5) .. (1,0)
               .. controls (1,-.5) and (1.8,-.5) .. (1.8,-1.2)
               .. controls (1.8,-1.5) and (1.7,-2) .. (1,-2)
               .. controls (.3,-2) and (-0.3,-1) .. (-0.3,0);
\begin{scope}[xshift=3.6cm]
\shadedraw[ball color=blue,xscale=-1] (-0.3,0) .. controls (-0.3,1) and (.3,2) .. (1,2)
               .. controls (1.7,2) and (1.8,1.5) .. (1.8,1.2)
               .. controls (1.8,.5) and (1,.5) .. (1,0)
               .. controls (1,-.5) and (1.8,-.5) .. (1.8,-1.2)
               .. controls (1.8,-1.5) and (1.7,-2) .. (1,-2)
               .. controls (.3,-2) and (-0.3,-1) .. (-0.3,0);
\end{scope}
\draw[green, line width=0.5mm, xshift=2.5cm,yshift=1cm] (0,0)--(0.2,0.2);
\draw[green, line width=0.5mm, xshift=2.5cm,yshift=1cm] (0,0.2)--(0.2,0);
\end{scope}

\begin{scope}[xshift=8cm]
\node at (1.3,2) {fibre over $C_2$};
\begin{scope}[scale=0.7]
   \shadedraw[ball color=gray!30!white] (-0.3,0) .. controls (-0.3,1) and (.3,2) .. (1,2)
               .. controls (1.7,2) and (1.8,1.5) .. (1.8,1.2)
               .. controls (1.8,.5) and (1,.5) .. (1,0)
               .. controls (1,-.5) and (1.8,-.5) .. (1.8,-1.2)
               .. controls (1.8,-1.5) and (1.7,-2) .. (1,-2)
               .. controls (.3,-2) and (-0.3,-1) .. (-0.3,0);
\begin{scope}[xshift=3.6cm]
\shadedraw[ball color=gray,xscale=-1] (-0.3,0) .. controls (-0.3,1) and (.3,2) .. (1,2)
               .. controls (1.7,2) and (1.8,1.5) .. (1.8,1.2)
               .. controls (1.8,.5) and (1,.5) .. (1,0)
               .. controls (1,-.5) and (1.8,-.5) .. (1.8,-1.2)
               .. controls (1.8,-1.5) and (1.7,-2) .. (1,-2)
               .. controls (.3,-2) and (-0.3,-1) .. (-0.3,0);
\end{scope}
\draw[green, line width=0.5mm, xshift=1cm,yshift=1cm] (0,0)--(0.2,0.2);
\draw[green, line width=0.5mm, xshift=1cm,yshift=1cm] (0,0.2)--(0.2,0);
\draw[blue, line width=0.5mm, xshift=2.5cm,yshift=-1cm] (0,0)--(0.2,0.2);
\draw[blue, line width=0.5mm, xshift=2.5cm,yshift=-1cm] (0,0.2)--(0.2,0);
\end{scope}
\end{scope}

\end{tikzpicture}
\caption{The fibre structure over the singlet curves $C_1$ and $C_2$. Blue denotes the section $S$ and green the section $U$.}\label{fig:fibre}
 \end{figure}

Massless matter charged under the gauge group $U(1)$ arises at those loci on ${\cal B}$ over which the fibre splits into two components intersecting as the affine Dynkin diagram of $SU(2)$.
There are two types of such splittings:
By inspection of \eqref{eq:Bl1124} one finds that over the codimension-two locus
\begin{eqnarray} \label{eq:C1def}   
C_1: \, (b_2, c_3)
\end{eqnarray}
the fibre factorises into
\begin{equation} \label{eq:PC1}
{P_1}|_{C_1} = s \, (w^2 + b_{0} s\, u^2 \, w + b_1 \, u\, v\, w + c_{0}\, s^2\, u^4 + c_{1}\, s\,  u^3 \, v + c_{2} \, u^2 \, v^2).
\end{equation}
The two factors define two rational curves  $\mathbb P^1_s$ and $\mathbb P^1_{\rm res.}$.
In particular, the rational section $S$ degenerates over $C_1$, where it wraps an entire fibre component $\mathbb P^1_s$. 
This type of behaviour had first been discussed in the $U(1)$ restricted Tate model, which is a special case of the hypersurface \eqref{eq:Bl1124}, in \cite{Grimm:2010ez,Braun:2011zm,Krause:2011xj}. 
M2-branes wrapping the fibre component $\mathbb P^1_{\rm res.}$ yield states with charge $q = \int_{\mathbb P^1_{\rm res.}} \tw = + 2$ since the section $S$ intersects $\mathbb P^1_{\rm res.}$ in two points and the section $U$ intersects the fibre only in one point on $\mathbb P^1_s$, cf.\ Figure~\ref{fig:fibre}. This will become important in section \ref{sec:the_field_theory_picture}, to which we refer for more details on the interpretation of M2-branes wrapped on the distinct fibre components.

As explained in \cite{Morrison:2012ei}, 
the fibre can also factorise in two components none of which is wrapped by the section $S$. This factorisation must persist in the singular blow-down of \eqref{eq:Bl1124} corresponding to $s\equiv 1$.
To find this locus one 
completes the square in $w$ and writes the hypersurface equation as
\begin{equation}
\begin{aligned}
P_1 = &\left[  w + \frac{1}{2}(b_{0}  u^2 + b_1  u v + b_2 v^2)\right]^2\\ 
&+ (c_{0}-\frac{1}{4}b_{0}^2) u^4 + (c_{1}-\frac{1}{2}b_{0}b_1) u^3v + (c_{2} - \frac{1}{2}b_{0}b_2-\frac{1}{4}b_1^2) u^2 v^2 + (c_3 - \frac{1}{2}b_1 b_2) u v^3 - \frac{1}{4}b_2^2 v^4.
\end{aligned}
\end{equation}
 The equation factorises if the polynomial in $ u$ and $v$ in the second line is a perfect square. An ansatz of the form $(A  u^2 + B uv + Cv^2)^2$ yields five equations, three of which determine the coefficients $A$, $B$ and $C$ in terms of the $b_i$ and $c_i$. Solving for $A$ and $B$ places $b_2$ and $2c_3 - b_1 b_2$ in the denominator and thus the factorization is valid away from the vanishing locus defined by $(b_2, 2c_3 - b_1b_2)$. The remaining two equations have the form
\begin{equation}\label{eq:poly6_singlet_loci}
\begin{aligned}
& - c_1 b_2^4 +b_1b_2^3 c_2+b_0 b_2^3 c_3-b_1^2 b_2^2 c_3-2 b_2^2c_2 c_3+3 b_1 b_2 c_3^2-2c_3^3 =0, \\
&-c_3^2 b_0^2 +b_1
   b_2 b_0^2 c_3-b_1 b_2^2 b_0
   c_1+b_2^2 c_1^2+b_1^2 b_2^2
   c_0-4 b_1 b_2 c_0 c_3+4 c_0
   c_3^2 =0.
\end{aligned}
\end{equation}
These are valid away from the ideal $(b_2, 2c_3 - b_1 b_2) = (b_2, c_3)$.  The two polynomials \eqref{eq:poly6_singlet_loci} generate an ideal of the polynomial ring $\mathbb C[b_i,c_i]$ over ${\cal B}$ whose vanishing locus defines a complicated affine variety on ${\cal B}$.
The individual irreducible components are found by decomposing the original ideal \eqref{eq:poly6_singlet_loci}  into its prime ideals and selecting only the codimension-two loci. 
This method was used in \cite{Cvetic:2013uta,Cvetic:2013jta} (see also \cite{Lin:2014qga}) to determine the irreducible singlet curves in an analogous fibration with Mordell-Weil group of rank two \cite{Borchmann:2013jwa,Cvetic:2013nia,Cvetic:2013uta,Borchmann:2013hta} (see \cite{Cvetic:2013qsa,Klevers:2014bqa} for an analysis in this spirit of, among other things, the singlet locus of even more general fibrations). \\

\noindent We analyse the locus \eqref{eq:poly6_singlet_loci} in this spirit with SINGULAR \cite{singular}. By taking the saturation of $C_1$ in \eqref{eq:poly6_singlet_loci} we find that it does not split into further components. The only component is
\begin{equation}\label{eq:singly_charged_C2}
C_2: \, \{\text{An ideal with 15 generators}\},
\end{equation}
which is checked to be of codimension two, and thus the generators do not intersect transversely. The prime ideal analysis furthermore shows that $C_2$ has a singular locus, a point of self-intersection to be discussed in more detail momentarily. \\

\noindent By construction, the fibre over $C_2$ splits into homologous $\mathbb P^1$s whose intersection numbers with the $U(1)$ generator imply that  at $C_2$ a matter state of charge $+1$ and its conjugate state are localised. For more details see again section \ref{sec:the_field_theory_picture}. 


\begin{figure}
\vspace{-0cm}
\centering \hspace{0cm}
\scalebox{1}{
\def\svgwidth{170pt} 
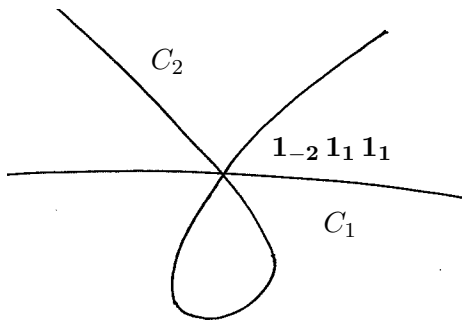 
}
\vspace{-0cm}
\caption{The singlet curves and the Yukawa coupling in codimension three.}
 \end{figure}

\noindent In codimension three on ${\cal B}$, i.e.\ at points, the singlet curves intersect. The singular locus of $C_2$ coincides with the intersection point $C_1 \cap C_2$ and is found, with the help of the prime decomposition technique, to be located at 
\begin{equation} \label{eq:C1capC2}
C_1 \cap C_2 :  (b_2, \, c_3, \,b_1^2 c_0 - b_0 b_1 c_1+b_0^2c_2+c_1^2-4c_0c_2)\, .
\end{equation}
The fibre over the singlet Yukawa point factors into three rational curves intersecting as the affine $SU(3)$ Dynkin diagram as seen from the hypersurface equation
\begin{equation}
P|_{C_1 \cap C_2} =  s \left( \tilde{w}^2  - \big(\frac{u}{\sqrt{b_1^2-4c_{2}}}[(c_{1} - \frac{1}{2}b_{0}b_1)su + 2(c_{2}-\frac{1}{4}b_1^2)v] \big)^2\right) \, .
\end{equation}
The first factor defines the curve $\mathbb P^1_s$, with $U(1)$ charge $-2$, and the second factor is the difference of two squares which defines two curves, each with charge $1$. This confirms the presence of a Yukawa coupling $\mathbf{1}_{-2}\, \mathbf{1}_1\, \mathbf{1}_1$ at this point.

\subsection{A torus-fibration with \texorpdfstring{$\mathbb Z_2$}{Z2} symmetry }\label{sec:Z2fibration}

As recalled in the previous section, the appearance of a rank-one Mordell-Weil group and  thus the 
presence of a $U(1)$ gauge group factor in F-theory fibrations of the form \eqref{eq:Bl1124} is a consequence of the non-generic form of the hypersurface equation in which a potential quartic term $v^4$ is set to zero. By starting with the blow-down of  $s$ in \eqref{eq:Bl1124}  and allowing for such a non-vanishing quartic term,
one arrives at the smooth genus-one fibration whose fibre is described by 
the hypersurface equation 
\begin{equation}\label{eq:hypersurface-Z2-model}
\begin{aligned}
P_2 =  w^2 + b_0 u^2 w  + b_1 u v w + b_2 v^2 w  + c_0 u^4  + c_1 u^3 v  + c_2 u^2 v^2  + c_3 u v^3  + c_4 v^4
\end{aligned}
\end{equation}
in $\mathbb{P}_{1,1,2}$. 
The scaling relations of $u$, $v$, $w$ as well as the classes of $b_i$ and $c_i$ are as in section \ref{sec:U(1)fibration}. 

The model was first discussed in \cite{Braun:2014oya,Morrison:2014era} as an example of a genus-one fibration with a bi-section (see also \cite{Anderson:2014yva}).
The two rational solutions ${\rm Sec}_1$ and ${\rm Sec_2}$ (cf.~\eqref{eq:sec01}) of the intersection of $u=0$ with the blow-down of \eqref{eq:Bl1124}
combine, as a consequence of $c_4 \neq 0$, into the two branches of a square root which are exchanged by a monodromy and thus cannot be distinct any more.
Indeed, 
\begin{equation}
\left.{P_2}\right|_{u=0} = \left(w + \frac{1}{2} \left(b_2 + \sqrt{b_2^2 - 4\,c_4}\right) \, v^2\right) \left(w + \frac{1}{2} \left(b_2 - \sqrt{b_2^2 - 4\,c_4}\right) \,  v^2\right). 
\end{equation}
Thus, the divisor $U: u=0$ describes a bi-section into which the zero-section and the rational section of  \eqref{eq:Bl1124} combine \cite{Braun:2014oya,Morrison:2014era}. 

The fibre continues to factorise in codimension two \cite{Braun:2014oya,Morrison:2014era}.
This locus has been derived in \cite{Morrison:2014era} in a fashion similar to the procedure for the fibration with $c_4=0$ reviewed around \eqref{eq:poly6_singlet_loci}.
However, the analysis of \cite{Morrison:2014era} is a priori valid for base spaces of complex dimension two such that we need to apply special care in deriving the analogous singlet locus for base spaces of dimension three.
To this end, we first rewrite $P_2$ by completing the square in $w$ as
\begin{eqnarray}
P_2 = \tilde w^2 +  a_0 u^4 + a_1 u^3 v + a_2 u^2 v^2 + a_3 u ^2 v^2 + a_4 v^4
\end{eqnarray}
with $\tilde w = w + \frac{1}{2} (b_0 u^2 + b_1 u v + b_2 v^2)$ and 

\begin{eqnarray}
a_0 &=& - c_0 + \frac{1}{4} b_0^2,  \qquad a_1 = - c_1 + \frac{1}{2} b_0 b_1, \\
a_2 &=&  - c_2 + \frac{1}{2} b_0 b_2 + \frac{1}{4} b_1^2,   \qquad a_3 = - c_3 + \frac{1}{2} b_1 b_2,  \qquad a_4 = -c_4 + \frac{1}{4} b_2^2.
\end{eqnarray}
There are now two cases to distinguish:
If $a_4 \neq 0$ we can make the ansatz \cite{Morrison:2014era}
\begin{eqnarray} \label{eq:factorisation1}
P_2 = \tilde w^2 - a_4 ( A u^2 + B u v + v^2 )^2 = (\tilde w - \sqrt{a_4} ( A u^2 + B uv + v^2)) (\tilde w + \sqrt{a_4} ( A u^2 + B uv + v^2)).
\end{eqnarray}
Comparing this ansatz to $P_2$  identifies the polynomials $A$ and $B$ as \cite{Morrison:2014era}
\begin{eqnarray}
A= \frac{4 a_2 a_4 - a_3^2}{8 a_4^2}, \qquad B = \frac{a_3}{ 2 a_4}
\end{eqnarray}
and in addition gives two constraints $p_1 = p_2 =0$  
with 
\begin{equation}
\begin{aligned} \label{eq:singlet_factors_poly4}
p_1=&b_{2}^{6} c_{0} -  b_{1}^{2} b_{2}^{3} c_{4} b_{0} + b_{1} b_{2}^{4}
c_{3} b_{0} -  b_{2}^{5} c_{2} b_{0} + b_{2}^{4} c_{4} b_{0}^{2} +
b_{1}^{4} c_{4}^{2} - 2 b_{1}^{3} b_{2} c_{4} c_{3} + b_{1}^{2}
b_{2}^{2} c_{3}^{2} + 2 b_{1}^{2} b_{2}^{2} c_{4} c_{2}
 +\\& - 2 b_{1}
b_{2}^{3} c_{3} c_{2} + b_{2}^{4} c_{2}^{2} - 12 b_{2}^{4} c_{4} c_{0} +
4 b_{1}^{2} b_{2} c_{4}^{2} b_{0} - 4 b_{1} b_{2}^{2} c_{4} c_{3} b_{0}
-  b_{2}^{3} c_{3}^{2} b_{0} + 8 b_{2}^{3} c_{4} c_{2} b_{0}
 +\\&- 8
b_{2}^{2} c_{4}^{2} b_{0}^{2} + 2 b_{1}^{2} c_{4} c_{3}^{2} - 2 b_{1}
b_{2} c_{3}^{3} - 8 b_{1}^{2} c_{4}^{2} c_{2} + 8 b_{1} b_{2} c_{4}
c_{3} c_{2} + 2 b_{2}^{2} c_{3}^{2} c_{2} - 8 b_{2}^{2} c_{4} c_{2}^{2}
 +\\&
+ 48 b_{2}^{2} c_{4}^{2} c_{0} + 4 b_{2} c_{4} c_{3}^{2} b_{0} - 16
b_{2} c_{4}^{2} c_{2} b_{0} + 16 c_{4}^{3} b_{0}^{2} + c_{3}^{4} - 8
c_{4} c_{3}^{2} c_{2} + 16 c_{4}^{2} c_{2}^{2} - 64 c_{4}^{3} c_{0}
\,,\\
p_2=&-\tfrac{1}{2} b_{1}^{3} b_{2} c_{4} + \tfrac{1}{2} b_{1}^{2} b_{2}^{2}
c_{3} - \tfrac{1}{2} b_{1} b_{2}^{3} c_{2} + \tfrac{1}{2} b_{2}^{4} c_{1}
+ b_{1} b_{2}^{2} c_{4} b_{0} - \frac{1}{2} b_{2}^{3} c_{3} b_{0} +
b_{1}^{2} c_{4} c_{3} 
 +\\&- \tfrac{3}{2} b_{1} b_{2} c_{3}^{2} + 2 b_{1}
b_{2} c_{4} c_{2} + b_{2}^{2} c_{3} c_{2} - 4 b_{2}^{2} c_{4} c_{1} - 4
b_{1} c_{4}^{2} b_{0} + 2 b_{2} c_{4} c_{3} b_{0} + c_{3}^{3} - 4 c_{4}
c_{3} c_{2} + 8 c_{4}^{2} c_{1}\,.
\end{aligned} 
\end{equation}
This shows that the factorisation \eqref{eq:factorisation1} occurs over the variety defined by
\begin{eqnarray}
\tilde C_1 = \{p_1 = 0\} \cap \{p_2 = 0\} - \{ a_4=0 \} \cap \{p_1 = 0\} \cap \{p_2 = 0\}.
\end{eqnarray}
Note that on $\{p_1 = 0\} \cap \{p_2 = 0\}$, $a_4=0$ implies $a_3=0$.
For base spaces of dimension two, as analysed in \cite{Morrison:2014era}, this is the complete $I_2$ locus.
For complex three-dimensional bases, by contrast, also the case $a_4 =0$ must be taken into account. 
For $a_4 =0$, the above ansatz for the factorisation is not valid, and a different ansatz must be chosen. Let us assume that $a_0 \neq 0$ and make the ansatz
\begin{eqnarray} \label{eq:factor2}
P_2|_{a_4=0} = \tilde w^2  - a_0 (u^2 + C u v )^2 = (\tilde w + \sqrt{a_0} u (u + D v))  (\tilde w - \sqrt{a_0} u (u + D v)) . 
\end{eqnarray}
Comparing with the original form of the hypersurface identifies
\begin{eqnarray}
D= \frac{a_1}{2\, a_0}
\end{eqnarray} 
and gives two more constraints $a_3 = a_1^2 - 4 a_2 a_0=0$. Note that for a generic base ${\cal B}$ and a generic choice of sections $b_i$, $c_i$, we have that $a_0 \neq 0$ at   $\{a_4 = 0\} \cap \{a_3 = 0\} \cap \{ a_1^2 - 4 a_2 a_0 = 0\}$.  Thus the above factorisation \eqref{eq:factor2} holds at the codimension-three loci
\begin{eqnarray}
\tilde C_2 = \{a_4 = 0\} \cap \{a_3 = 0\} \cap \{ a_1^2 - 4 a_2 a_0 = 0\}.
\end{eqnarray}
In all, the fibre factorises into an $I_2$-fibre at the locus
\begin{eqnarray} \label{eq:defC}
C= \tilde C_1 \cup \tilde C_2.
\end{eqnarray}
It turns out that this locus $C$ has a representation as the vanishing locus of a prime ideal. This ideal is defined as the saturation of the ideal generated by $a_3$ and $a_4$ within the ideal generated by $p_1$ and $p_2$.
Using the prime decomposition technique gives a representation of this ideal as being generated by 16 non-transversely intersecting polynomials.
One can show that the points $\tilde C_2$ all lie on the variety defined by this prime ideal and are indeed the only loci on $p_1 \cap p_2$ for which $a_4=0$.
 Importantly, we can use our representation of $C$ as a prime ideal to show that $C$ defines an irreducible, smooth curve on ${\cal B}$. This is the locus where the fibre is of $I_2$-type.

M2-branes wrapping either of the two fibre components over $C$ give rise, in the F-theory limit, to massless singlet states. Due to the absence of  sections, it is not possible to define a $U(1)$ generator $\tw$ as in equation \eqref{eq:Shiodamap}. 
As explained above, $S$ and the zero-section combine into the bi-section $U$ which intersects the fibre class in two points. Global monodromy effects interchange these two points over the base ${\cal B}$ \cite{Braun:2014oya}.
Nonetheless it is possible to define the notion of a $\mathbb Z_2$-charge of the singlet states with respect to the section $U$, as we now discuss:
By plugging $u=0$ into \eqref{eq:singlet_factors_poly4} one confirms that the divisor $U$ intersects each of the two split fibre components over the locus $\tilde C_1$ in a single point, given, respectively, by
\begin{eqnarray}
u=0, \qquad  w = \pm \frac{1}{\sqrt{b_2^2 - 4 c_4}} v^2.
\end{eqnarray} 
Again, since $b_2^2 - 4 c_4 \neq 0$ along $\tilde C_1$ these two points are well-defined. 
However, as we approach the points $\tilde C_2$, the two intersection points coincide not only with each other but also with one of the intersection points of the rational lines of the $I_2$ fibre. This is the expected behaviour of the  monodromy of the two points $u=0$ around this codimension-three locus on ${\cal B}$. As pointed out at the very end of section 6 in \cite{Braun:2014oya}, such a behaviour prevents us from defining a well-defined $U(1)$ charge, because if we would take the differences of the  two points the charges would flip sign when going around the monodromy. However if we take the sum of the points this problem does not appear any more. In order for the two states wrapping the two different $\mathbb P^1$s of the $I_2$ fibre to carry conjugate charges we can only take the charges modulo two.  
The interpretation of this behaviour is that the bi-section $U$ generates a $\mathbb Z_2$ quantum number when intersected with the rational lines of the fibre. 
The consequences of this quantum number as a discrete selection rule will be apparent when engineering extra gauge symmetry and hence massless matter states, cf.~section~\ref{sec:the-SU(5)xZ2-case}.

\subsection{Compensating \texorpdfstring{$G_4$}{G4}-flux in 4D conifold transitions}\label{sec:compensator_G4-flux}

The transition between the two elliptic fibrations \eqref{eq:Bl1124} and \eqref{eq:hypersurface-Z2-model} is a deformation of the blown-down model with 
$c_4=0$ to $c_4 \neq 0$. Let us denote by $\hat Y_4$ the smooth Calabi-Yau fourfold described by $P_1$ in \eqref{eq:Bl1124} and by $X_4$ the smooth Calabi-Yau fourfold described by $P_2$ in \eqref{eq:hypersurface-Z2-model}.
After the transition, the analogue of the curve $C_1$---see \eqref{eq:C1def}---of states $\mathbf{1}_{2}$ in $\hat Y_4$ ceases to correspond to a charged matter locus on $X_4$, while the locus $C_2$ of $\mathbf{1}_{1}$ states on $\hat Y_4$ gets deformed exactly into the $I_2$ locus $C$ on $X_4$, defined in \eqref{eq:defC}. The singular points on $C_2$ in $\hat Y_4$ given by \eqref{eq:C1capC2} are smoothed out, and $C$ on $X_4$ is indeed a smooth irreducible curve. 
This picture is consistent with an interpretation as a four-dimensional Higgsing by giving a D-flat VEV to the doubly charged singlets along $C_1$, and would, in Type IIB language, correspond to a process of brane recombination. 
The Higgsing breaks $U(1) \rightarrow \mathbb Z_2$ \cite{Morrison:2014era}, and all remaining states are charged only under the discrete $\mathbb Z_2$. Further evidence for this picture will be provided later when introducing extra non-abelian gauge groups and analysing the structure of their Yukawa interactions. 

Mathematically, the deformation corresponds to a conifold transition as has been worked out already in the six-dimensional context in \cite{Morrison:2014era} (see also \cite{Anderson:2014yva}). 
We now discuss the consequences of this deformation in more detail by focusing again on generic base spaces ${\cal B}$ of complex dimension three, where interesting novel features occur compared to the six-dimensional context which are related to the appearance of $G_4$-flux.

By direct comparison one finds, via repeated use of the adjunction formula, that the Euler characteristics before and after the deformation are related as 
\begin{eqnarray}
\chi(\hat Y_4) = \chi(X_4) + 3 \chi(C_1), 
\end{eqnarray}
where
\begin{eqnarray}
\chi(C_1) =  \int_{C_1} c_1(C_1) =  - \int_{\cal B} [b_2] \wedge [c_3] \wedge [c_4]
\end{eqnarray}
denotes the Euler characteristic of the curve $C_1$ along which the Higgsing is performed. 
In four-dimensional F-theory compactifications on $\hat Y_4$, $\chi(\hat Y_4)/24$ is known to count the curvature induced D3-charge, and in presence of  $G_4$-flux, the number of D3-branes $n_{D3}$ is given via the tadpole cancellation condition by
\begin{eqnarray}
n_{D3} = \tfrac{1}{24}\chi(\hat Y_4) - \tfrac{1}{2} \int_{\hat Y_4} \hat G_4 \wedge \hat G_4,
\end{eqnarray}
where the hat in $\hat G_4$ reminds us that this flux is defined on the fourfold $\hat Y_4$.
The smooth deformation in $c_4$ cannot lead to a jump in this topological quantity. Indeed, it is well-known \cite{Gaiotto:2005rp,Collinucci:2008pf} that brane recombinations lead to a change in the gauge flux in precisely such a way as to cancel the difference of the curvature induced D3-brane charges such that $n_{D3}$ does not change. 
This has been analysed in \cite{Braun:2011zm,Krause:2012yh} in the context of the $U(1)$ restricted Tate model, which is a special case of the present example in which no $\mathbf{1}_{2}$ states are present and thus the Higgsing is from $U(1) \rightarrow \emptyset$, as well as in a general M-theory context in \cite{Intriligator:2012ue}.

Let us first assume, for simplicity, that $ \hat G_4=0$ on $\hat Y_4$ before the transition. Then by the above arguments, a canonical gauge flux must appear on $X_4$ such that
\begin{eqnarray} \label{eq:tadmatch1}
\tfrac{1}{2}  \int_{X_4}  \, G_4 \wedge G_4 =  - \tfrac{1}{8} \chi (C_1).
\end{eqnarray}

We now make a proposal for the form of a canonical such $G_4$ on $X_4$, inspired by the analysis for the $U(1)$ restricted model in \cite{Braun:2011zm} and consistent with the results of \cite{Intriligator:2012ue}. 
On a specific locus in the complex structure moduli space of $X_4$ on which
\begin{eqnarray}
c_4 = \rho \, \tau,
\end{eqnarray}
new algebraic 4-cycles appear. These can be written as a complete intersection on the ambient space $X_5$, into which $X_4$ is embedded as a hypersurface, as 
\begin{eqnarray}
\sigma_0 &=& \{ u = 0 \} \cap \{ w = 0 \}  \cap \{\rho = 0 \}, \\
\sigma_1 &=& \{ u = 0 \} \cap \{ w = - b_2 v^2 \}  \cap \{\rho = 0 \}. 
\end{eqnarray}
Our proposal for $G_4$ is now to take the combination
\begin{eqnarray} \label{eq:P-flux}
G_4(P) = [\sigma_1] - \frac{1}{2} \, U \wedge P,
\end{eqnarray}
in terms of $U$, the bi-section on $X_4$, and the divisor $P: \{\rho =0\}$. 
This flux is an element of $H^{2,2}(X_4)$ precisely if $c_4 = \rho \,  \tau$. 

First, we observe that this flux satisfies the two conditions
\begin{eqnarray} \label{eq:trans}
\int_{X_4} G_4(P) \wedge \pi^{-1}D_a \wedge \pi^{-1}D_b = 0,   \qquad \int_{X_4} G_4(P) \wedge U \wedge \pi^{-1}D_b = 0 \qquad \forall D_a, D_b \in H^{1,1}({\cal B}).
\end{eqnarray}
These can be viewed as analogous to the transversality conditions with respect to the zero-section in elliptic fibrations. 
Second, the flux induced tadpole can be computed as 
\begin{eqnarray} \label{eq:fluxTAD}
\tfrac{1}{2} \int_{X_4} G_4(P) \wedge G_4(P) = \int_{{\cal B}} P \wedge \left( -\tfrac{1}{4 }P \wedge [c_3] + \tfrac{1}{2} [c_3] \wedge [b_2] \right).
\end{eqnarray}
For the specific choice
\begin{eqnarray}
P = \frac{1}{2} [c_4] \equiv  - \frac{1}{2} c_1(T_{C_1})
\end{eqnarray}
this indeed reduces to
\begin{eqnarray}
\left.\tfrac{1}{2} \int_{X_4} G_4(P) \wedge G_4(P)\right|_{P = \frac{1}{2} [c_4]} =  - \tfrac{1}{8} \chi (C_1),
\end{eqnarray}
where we have used the fact that $[c_4] = 2 [b_2]$, see Table~\ref{tab:coeff1}.
Therefore this flux precisely cancels the change of the curvature induced D3-brane charge as it must. 
Let us stress that the flux $G_4(P)$ induces a superpotential on $X_4$ whose critical locus is precisely given by
the locus on which $c_4 = \rho \tau$ such that $G_4$ is of $(2,2)$ type \cite{Braun:2011zm}. Finally note that the flux can  be shown to obey the quantisation condition $G_4(P) + \frac{1}{2}c_2(X_4) \in H^{2,2}(X_4,\mathbb Z)$. 

The derivation of these results repeatedly employs, as usual, the homological relations between the toric divisors $u$, $w$ and $v$ on the toric ambient space into which $X_4$ is embedded as a hypersurface of class $2 [w]$. This way self-intersections can be reduced to transverse intersections, which can explicitly be evaluated. 
For instance,
\begin{eqnarray}
&& \int _{X_4} \sigma_1 \wedge \pi^{-1}D_a \wedge \pi^{-1}D_b = \int_{X_5} [u] \wedge [w] \wedge P \wedge \pi^{-1}D_a \wedge \pi^{-1}D_b = \int_{\cal B} P \wedge \pi^{-1}D_a \wedge \pi^{-1}D_b, \\
&& \int _{X_4} U \wedge P \wedge \pi^{-1}D_a \wedge \pi^{-1}D_b = \int_{X_5} [u] \wedge [w^2] \wedge P \wedge \pi^{-1}D_a \wedge \pi^{-1}D_b = 
2 \int_{\cal B} P \wedge \pi^{-1}D_a \wedge \pi^{-1}D_b, \nonumber
\end{eqnarray}
which shows the first transversality condition. 
For the second equation in \eqref{eq:trans} as well as for the D3-tadpole, self-intersections are required, e.g.
\begin{eqnarray}
\int_{X_4} \sigma_1 \wedge U \wedge P = \int_{X_5} [u] \wedge [w] \wedge P \wedge [u] \wedge P = - \int_{\cal B} (\beta - \bar {\cal K}) \wedge P \wedge P, 
\end{eqnarray}
where we have used that $[u] \wedge [u] = [u] \wedge ([v] - (\beta - \bar {\cal K}))$ as can be read off from Table~\ref{tab:scaling} together with the fact that $u\, v \, w$ is in the SR-ideal of $X_4$. 
For the self-intersection of $\sigma_1$ we use, inspired again by \cite{Braun:2011zm}, that
\begin{eqnarray}
\int_{X_4} [\sigma_1] \wedge [\sigma_1] = \int_{\sigma_1} c_2(N_{\sigma_1 \subset X_4}) = \int_{\sigma_1}  \frac{c_2(N_{\sigma_1 \subset X_5})}{c_2(N_{X_4 \subset X_5})}.
\end{eqnarray}
This can be evaluated by expanding
\begin{eqnarray}
c(N_{\sigma_1 \subset X_4}) = \frac{(1 + [u])(1+ [w]) (1+ [\rho]) }{1 + 2 [w]}
\end{eqnarray}
and repeatedly using relations of the form  $[u] \wedge [u] = - [u]\wedge \alpha$ and $[w]\wedge[w] = w \wedge (\beta-2 \alpha)$. This gives
\begin{eqnarray}
\int_{X_4} [\sigma_1] \wedge [\sigma_1]  = \int_{\cal B} P \wedge P \wedge (-\bar{\cal K}) + \int_{\cal B} P \wedge ( 3 \bar{\cal K} -\beta) \wedge (2 \bar{\cal K} - \beta)
\end{eqnarray}
and eventually \eqref{eq:fluxTAD}.

The above results can be generalized and further interpreted as follows:
Suppose that before the transition, on the fourfold $\hat Y_4$, we have switched on non-trivial $U(1)$ gauge flux of the form
\begin{eqnarray}
\hat G_4({F}) = \tw \wedge \pi^{-1}{ F} = (S - U - {\bar {\cal K}} - [b_2]) \wedge \pi^{-1} F, \qquad \quad F \in H^{1,1}({\cal B}).
\end{eqnarray}
Then in order to compensate for the change of the Euler characteristic of the fourfold in the transition, the previous expression \eqref{eq:tadmatch1} must be generalized to
\begin{eqnarray}
\frac{1}{2}  \int_{X_4}  \, G_4( P ) \wedge G_4( P ) =  - \frac{1}{8} \chi (C_1) + \frac{1}{2} \int_{\hat Y_4} \hat G_4(F) \wedge \hat G_4(F) .
\end{eqnarray}

With the help of the homological relations of Tables~\ref{tab:scaling} and \ref{tab:coeff1} on the toric ambient space of $\hat Y_4$ it is possible to derive the expression
\begin{eqnarray}
 \frac{1}{2} \int_{\hat Y_4} \hat G_4(F) \wedge \hat G_4(F)  = -  \int_{\cal B} F \wedge F \wedge (3 {\bar{\cal K}}- \beta) = - \int_{\cal B} F \wedge F \wedge [c_3].
\end{eqnarray}
Together with \eqref{eq:fluxTAD} one eventually finds that this constrains the class $P$ appearing in the flux \eqref{eq:P-flux} on $X_4$ after the transition to be
\begin{eqnarray} \label{eq:PFrel}
P = 2 F + \frac{1}{2} [c_4].
\end{eqnarray}
This result is again of a similar form as for the Higgsing or conifold transition of the $U(1)$ restricted Tate model \cite{Braun:2011zm}, except for the factor of $2$ appearing in front of $F$ in \eqref{eq:PFrel}. This is a consequence of the fact that, in the present situation, the Higgs field has $U(1)$ charge $2$, while for the $U(1)$ restricted model, the singlet which has be Higgsed had charge $1$. 
From the analysis in \cite{Krause:2011xj} (see also \cite{Intriligator:2012ue}), we see that the appearance of the charge of Higgs field in front of $F$ is indeed a beautiful consistency check of the entire setup: 
Since $P$ is the class of $\rho$ in $c_4 = \rho \, \tau$, we know that  $0 \leq P \leq [c_4]$ and thus, in view of \eqref{eq:PFrel},
\begin{eqnarray}
-\frac{1}{2} [c_4] \leq 2 F \leq \frac{1}{2} [c_4].
\end{eqnarray}
Integration over the Higgs curve $C_1$  implies the inequality
\begin{eqnarray} \label{eq:stablecondo}
-\frac{1}{2} c_1({\cal K}_{C_1}) \leq 2 c_1(L) \leq \frac{1}{2} c_1({\cal K}_{C_1}),
\end{eqnarray}
where $L$ is the line bundle on $C_1$ with first Chern class $c_1(L) = F|_{C_1}$ and we recall that $[c_4] = -c_1(T_{C_1}) = c_1({\cal K}_{C_1})$. 
The line bundle $L$ counts the ${\cal N}=1$ chiral multiplets on $C_1$, more precisely $h^0(C,(L^2 \otimes \sqrt{{\cal K}_{C_1}})$ counts chiral multiplets ${\bf 1}_{2}$ and 
$h^1(C,(L^2 \otimes \sqrt{{\cal K}_{C_1}})$ counts chiral multiplets of conjugate charge ${\bf 1}_{-2}$, and the power of $2$ is again due to the charge $2$ (see \cite{Bies:2014sra} for more details on the counting of vector-like zero modes in F-theory). A D-flat Higgsing with ${\bf 1}_{2} + c.c.$ is possible only if both $h^0(C,(L^2 \otimes \sqrt{{\cal K}_{C_1}})$  and $h^1(C,(L^2 \otimes \sqrt{{\cal K}_{C_1}})$ are non-zero such that massless vector-like pairs of Higgs fields exist in four dimensions. A necessary condition for this is \eqref{eq:stablecondo} \cite{Krause:2011xj}, and it is gratifying to see that this condition in turn follows from the geometric setup.

\subsection{The field theory picture}\label{sec:the_field_theory_picture}

In this section, we discuss the interpretation of the geometry from a field theory perspective. It is important to note that some aspects of the results we will obtain have already been mentioned in \cite{Morrison:2014era}, but we will add significant details to this picture. The M-theory geometry that we discuss can be understood in terms of the four-dimensional field theory associated to the F-theory compactification. More precisely the map from the geometry studied should be to a four-dimensional gauge theory on three-dimensional Minkowski space times a circle $M_{3}\times S^1$. Let us describe this theory from a three-dimensional perspective, keeping some of the Kaluza-Klein (KK) modes associated to the $S^1$ so that this is not a full dimensional reduction. The modes which are of interest for us are 2 three-dimensional $U(1)$ gauge fields $U(1)_0$ and $U(1)_1$. The $U(1)_0$ is associated to the component of the metric with one leg along the $S^1$, the so-called KK $U(1)$. The $U(1)_1$ is associated to the zero-mode (in terms of the KK tower) of the four-dimensional $U(1)$. There are also two three-dimensional scalars associated to these $U(1)$s, which we call $\xi_0$ and $\xi_1$. The first is just the component of the metric determining the size of the $S^1$, which we set to $\left<\xi_0\right>=1/R$ where $R$ is the radius of the $S^1$, and the second is the component of the four-dimensional $U(1)$ gauge field along the $S^1$. Importantly, the scalar $\xi_1$ is associated to the Wilson line of the four-dimensional gauge field along the $S^1$, $\xi_1 = \int_{S^1} A^1_{4D}$,
and the theory must be invariant under a shift $\xi_1 \rightarrow \xi_1 + \frac{1}{R}$, which corresponds to a large gauge transformation.


Finally, we introduce two four-dimensional fields charged under the four-dimensional $U(1)$ gauge field, $\Psi^1$ and $\Psi^2$, with the superscripts denoting the four-dimensional $U(1)$ charges. These fields lead to a KK tower of three-dimensional fields
\begin{eqnarray}
\Psi^i_n = \sum_{n=-\infty}^{n=+\infty} \psi^i_n e^{2 \pi i n y},
\end{eqnarray}
where $y$ is the coordinate along the $S^1$.  We are interested in the mass spectrum of these KK states of the charged fields. In the absence of a VEV for the Wilson line field $\xi_1$ their mass is just given by $m_n=\left|\frac{n}{R}\right|$. This gives a massless zero mode and a KK tower, each mass level of which comprises two KK modes corresponding to $\pm n$. However if we turn on a VEV for $\xi_1$, i.e.\ move on the Coulomb branch, their masses will change. Specifically, reducing the four-dimensional covariant derivative leads to the mass
\begin{eqnarray}
m^q_n = \left| q \left< \xi_1 \right> + \frac{n}{R} \right| \;, \label{eq:kkmass} 
\end{eqnarray}
where $q$ is the charge of the state under the four-dimensional $U(1)$. Indeed the two terms in \eqref{eq:kkmass} are of the same nature, with $R^{-1}$ being the VEV for $\xi_0$ and $n$ being the charge of the state under the KK $U(1)$. Now, importantly, the mass manifests the property of invariance under $\xi_1 \rightarrow \xi_1 + \frac{1}{R}$. Such a four-dimensional gauge transformation amounts to a re-arranging of the KK tower. Specifically as we turn on a VEV for $\xi_1$ the zero mode will gain a mass, while the pair of KK modes at the first excited level will split, one becoming lighter and the other heavier. As we reach $\xi_1=\frac{1}{R}$ one of the pair of KK modes will become massless, and the whole KK tower will look the same again.\footnote{Note that for a field of charge $q$ it is sufficient to shift $\xi_1 \rightarrow \xi_1 + \frac{1}{qR}$. One consequence of this is that, if we have also a state of charge one say, then there are $q$ different vacua for the charge one state which are all equivalent from the charge $q$ state perspective. In particular in a background where the state of charge $q$ obtains a VEV there are $q$ different vacua for the charge one state. This can also be thought of in terms of the spectrum of Wilson line operators in the four-dimensional theory that are not gauge equivalent, see for example \cite{Banks:2010zn}.} 

Let us now see how this picture is manifest in the geometry. The states $\psi_n^1$ and $\psi_n^2$ are associated to membranes wrapping certain components of the fibre. Specifically, consider the resolved $\textmd{Bl}^1{\mathbb P}_{1,1,2}$ model \eqref{eq:Bl1124} with two sections $U$ and $S$. 
Each section will lead to a $U(1)$ in three dimensions which we denote by $U(1)_U$ and $U(1)_S$. The identification with the field theory $U(1)$'s amounts to the Shioda map and reads
\begin{eqnarray}
U(1)_0 = U(1)_U \,,\qquad U(1)_1 = U(1)_S - U(1)_U\;.
\end{eqnarray}
The charges of the singlets just arise from integrating the M2 action over the components of the fibre that they wrap. 
As discussed in the previous chapter, the intersection of $U$ and $S$ with the fibre components, which we call component $A_i$ and $B_i$, is different.
Over the double-charged locus $C_1$, $A_2$ is the component wrapped by the section $S$, which is also the component intersected by $U$ in a point, while over the single-charged locus $C_2$, $A_1$ denotes the component intersected by $U$ and $B_1$ the component intersected by $S$. This situation is depicted in Figure~\ref{fig:fibre}.
The charges of the states wrapping them are simply given by the appropriate intersection numbers with the sections. In terms of $U(1)_0$ and $U(1)_1$ this gives the charges of
\begin{equation}
\begin{aligned}
Q\left(M_{A_2}\right) &= \left(+1,-2\right),\\
Q\left(M_{B_2}\right) &= \left(0,2\right),  \\
Q\left(M_{A_1}\right) &= \left(+1,-1\right),  \\
Q\left(M_{B_1}\right) &= \left(0,+1\right) \;.
\end{aligned}
\end{equation}
Here, for instance, $M_{A_1}$ denotes the state wrapping the $A_1$-cycle over the curve $C_2$ with the single-charged states. Recall that for each of the states there will also be anti-M2 states of opposite charge wrapping the same fibre components which will form their (four-dimensional $N=2$) superpartners, and in particular the Higgsing is along the D-flat direction where they have equal vacuum expectation values. 

Now the Higgsing, or deformation, of the Bl$^1\mathbb{P}_{1,1,2}$  model with two sections corresponds to giving a VEV to a state on the doubly-charged curve. 
Importantly, the deformation occurs \emph{after} first blowing down the divisor $S$, which corresponds to shrinking the B cycle over the double-charged locus.
Therefore, the Higgs must be the massless state after the blow-down, i.e.\ it is the state $M_{A_2}$. We see that this has charge 1 under $U(1)_0$ and therefore it is a first excited KK state. The fact that the Higgsing is of a KK state was first shown in \cite{Anderson:2014yva} and in particular this was shown to recover the appropriate Chern-Simons terms. For more general discussions of the importance of Kalzua-Klein modes in F/M-theory duality see also \cite{Cvetic:2013nia,Grimm:2013oga}.
From the charges we can also read off the remaining massless combination of three-dimensional $U(1)$'s which remains after Higgsing, it is
\begin{eqnarray}
U(1)_{\rm massless} =  2U(1)_0+U(1)_1.
\end{eqnarray}
 Indeed this combination is precisely the one found based on geometric data for explicit examples of manifolds in \cite{Anderson:2014yva} by performing a conifold transition. With our field-theoretic understanding  it is clear why the massless $U(1)$ in three dimensions must be given by this specific combination, and why it is
 the same combination that appears geometrically in all the examples of \cite{Anderson:2014yva} as it is just a property of the fibre which is independent of the base.

The fact that the Higgsed state, which must be massless since it corresponds to a deformation mode of the geometry, is a KK mode may seem a bit puzzling at first sight. But it is in perfect match with the field theory discussion we have presented as long as the VEV of $\xi_1$ is chosen to be $\left\langle \xi_1 \right\rangle=\frac{1}{2R}$, in which case the mass \eqref{eq:kkmass} vanishes. The fixing of the point along the Coulomb branch of $\xi_1$ forces us to resolve the single charged curve locus, i.e.\ since both M2 states wrapping the fibre components $A_1$ and $B_1$ over that locus are charged under $U(1)_1$ they must be massive and so the fibre components must have finite size. This is in perfect agreement with the fact that in the ${\mathbb P}_{1,1,2}$ fibration the single-charged locus is already resolved. Indeed, the mass of the two states is seen to be equal for this specific VEV of $\left\langle\xi_1\right\rangle$, both are at $M=1/2R$.  This implies that the area of the two fibre components is equal, as required by the presence of a multi-section, precisely as pointed out in \cite{Morrison:2014era}.

\begin{figure}
    \centering
\begin{tikzpicture}[>=stealth]
\node at (1.3,2) {Resolved quartic};
\begin{scope}[scale=.7]
   \shadedraw[ball color=gray!30!white] (-0.3,0) .. controls (-0.3,1) and (.3,2) .. (1,2)
               .. controls (1.7,2) and (1.8,1.5) .. (1.8,1.2)
               .. controls (1.8,.5) and (1,.5) .. (1,0)
               .. controls (1,-.5) and (1.8,-.5) .. (1.8,-1.2)
               .. controls (1.8,-1.5) and (1.7,-2) .. (1,-2)
               .. controls (.3,-2) and (-0.3,-1) .. (-0.3,0);
\begin{scope}[xshift=3.6cm]
\shadedraw[ball color=blue,xscale=-1] (-0.3,0) .. controls (-0.3,1) and (.3,2) .. (1,2)
               .. controls (1.7,2) and (1.8,1.5) .. (1.8,1.2)
               .. controls (1.8,.5) and (1,.5) .. (1,0)
               .. controls (1,-.5) and (1.8,-.5) .. (1.8,-1.2)
               .. controls (1.8,-1.5) and (1.7,-2) .. (1,-2)
               .. controls (.3,-2) and (-0.3,-1) .. (-0.3,0);
\end{scope}
\draw[green, line width=0.5mm, xshift=2.5cm,yshift=1cm] (0,0)--(0.2,0.2);
\draw[green, line width=0.5mm, xshift=2.5cm,yshift=1cm] (0,0.2)--(0.2,0);
\end{scope}
\draw[->] (4,0)--(7,0) node [midway,above] {blow down} 
                node [midway,below] {to sing.\ WS};
\begin{scope}[xshift=10cm]
\node at (-1.1,2) {Singular Weierstra\ss{}};
\begin{scope}[scale=.8]
\begin{scope}[xscale=-0.8,yscale=1.2]
   \shadedraw[ball color=blue] (0,0) to [out=90,in=170] (1.8,1.5) 
    to [out=-10,in=125] (2.8,0.8) to [out=-55,in=80] (3,0)
    to [out=120,in=0] (1.8,0.5) to [out=180,in=90] (1.3,0)
    to [out=-90,in=180] (1.8,-0.5) to [out=0,in=-120] (3,0)
    to [out=-80,in=55] (2.8,-0.8) to [out=-125,in=10] (1.8,-1.5)
    to [out=-170,in=-90] (0,0);
\end{scope}
\draw[green, line width=0.5mm, xshift=-1cm,yshift=1cm] (0,0)--(0.2,0.2);
\draw[green, line width=0.5mm, xshift=-1cm,yshift=1cm] (0,0.2)--(0.2,0);
\end{scope}
\end{scope}
\draw[->] (1.3,-2.)--(1.3,-4.5) node [midway,above,rotate=90] {blow down} 
                node [midway,below,rotate=90] {to sing.\ quartic};
\begin{scope}[xshift=0.4cm,yshift=-7cm]
\node at (.9,2.) {Singular quartic};
\begin{scope}[scale=.8]
\begin{scope}[xscale=0.8,yscale=1.2]
   \shadedraw[ball color=gray!30!white] (0,0) to [out=90,in=170] (1.8,1.5) 
    to [out=-10,in=125] (2.8,0.8) to [out=-55,in=80] (3,0)
    to [out=120,in=0] (1.8,0.5) to [out=180,in=90] (1.3,0)
    to [out=-90,in=180] (1.8,-0.5) to [out=0,in=-120] (3,0)
    to [out=-80,in=55] (2.8,-0.8) to [out=-125,in=10] (1.8,-1.5)
    to [out=-170,in=-90] (0,0);
\draw[green, line width=0.5mm, xshift=2.9cm,yshift=-.1cm] (0,0)--(0.2,0.2);
\draw[green, line width=0.5mm, xshift=2.9cm,yshift=-.1cm] (0,0.2)--(0.2,0);
\draw[blue, line width=0.5mm, xshift=2.9cm,yshift=-.1cm, rotate=30] (0,0)--(0.2,0.2);
\draw[blue, line width=0.5mm, xshift=2.9cm,yshift=-.1cm, rotate=30] (0,0.2)--(0.2,0);
\end{scope}
\end{scope}
\end{scope}

\end{tikzpicture}

\caption{The fibre over the charge-two locus $C_1$. As in Figure~\ref{fig:fibre}, blue denotes the section $S$ and green the section $U$.}\label{fig:fibre-degeneration_over_2charge_locus}
\end{figure}
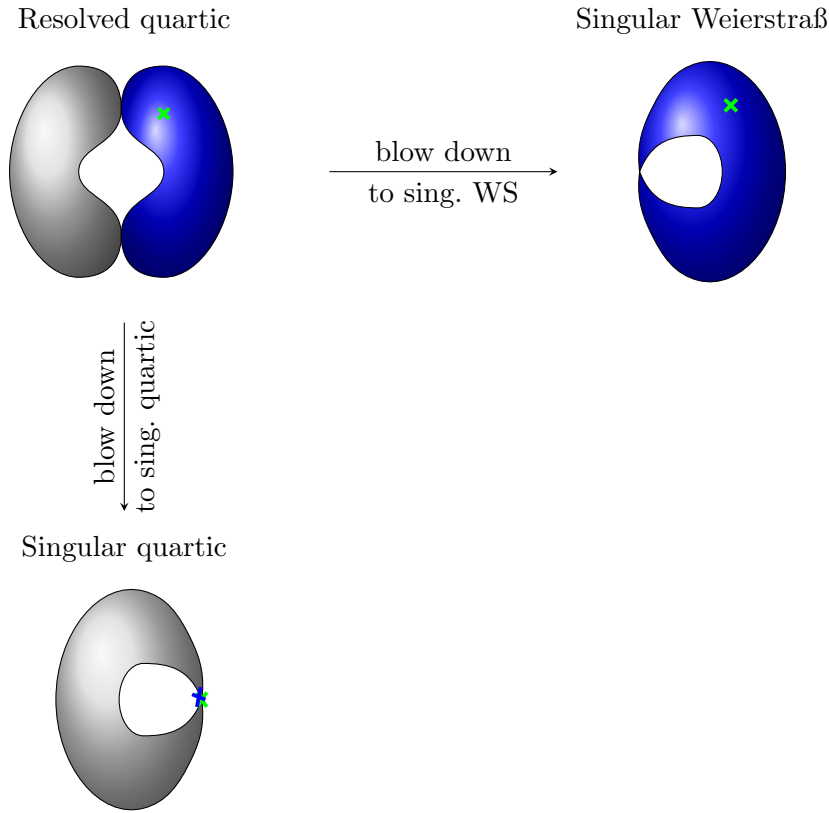

Since the Higgs state has charge 2 we expect two vacua in the three-dimensional theory after the Higgsing. We have outlined in detail one of the vacua and the associated geometry. To see the other vacuum we should consider the Jacobian or Weierstrass form of the fibration. Before the Higgsing this geometry is singular on both the charge-one locus $C_2$ and the charge-two locus $C_1$. To study this vacuum we wish to identify which of the M2 states is massless. For the singly-charged locus it is simple to identify the $B_1$-component of the fibre as shrinking to zero size, since the zero section does not intersect the pinch-point of the fibre.\footnote{In fact if we turn off the doubly-charged locus $C_1$ we reach the $U(1)$-restricted model studied in \cite{Grimm:2010ez,Krause:2012yh}, in which the singly-charged locus has been analysed in detail.} For the doubly charged locus, as shown in \cite{Morrison:2012ei}, it is possible to map the Weierstrass geometry to Bl$^1{\mathbb P}_{1,1,2}$ again; crucially, this time the map is to a different blow-down of the doubly-charged locus where the $B_2$-component vanishes. The difference between the two blow-downs is depicted in Figure~\ref{fig:fibre-degeneration_over_2charge_locus}.
We can therefore identify the massless states as coming from $M_{B_2}$ and $M_{B_1}$. These states are massless only at  $\left\langle\xi_1\right\rangle=0$, and so giving a VEV to the Higgs state $M_{B_2}$ fixes us to that point on the Coulomb branch with $M_{B_1}$ massless, which implies that it is not possible to resolve the singly charged locus. This is in perfect agreement with the fact that the Jacobian fibration of the deformed geometry does not admit a K\"ahler resolution \cite{Braun:2014oya}. It is interesting to note that in the two vacua, or geometrically in the ${\mathbb P}_{1,1,2}$ fibration and its Jacobian, the Higgs is in fact a different field. In one case it is $M_{B_2}$ and in the other case $M_{A_2}$, which, importantly, have different KK number. 

These considerations offer an interesting perspective on the multiple vacua one expects from a four-dimensional Higgsed theory on a circle: they are different Higgs branches in the three-dimensional theory for different KK modes of the four-dimensional field. It would be interesting to study this correspondence further (and its relation to the line operator spectrum of the theory), both from a field theory perspective and from the geometry side. Indeed we expect that the general picture is one where the $n$ vacua coming from a four-dimensional Higgs field of charge $n$ are associated to Higgsing $n$ different KK modes. Similarly from a geometric point of view the existence of $n$ vacua should correspond to $n$ different blow-downs. Finally it is important to note that these vacua are really physically distinct and do not correspond to a different formulation of the same physics.

\subsection{The 3D and 4D gauge group}

Although in the F-theory limit the four-dimensional gauge group after the Higgsing is $\mathbb{Z}_2$, the three-dimensional gauge groups arising from the physically distinct Higgsings in the ${\mathbb P}_{1,1,2}$ fibration and its Jacobian differ. Consider first the Jacobian, here the state which is being Higgsed has charges $\left(0,2\right)$ and therefore breaks directly in three dimensions the gauge group as $U(1)_0\times U(1)_1 \rightarrow U(1)_0 \times \mathbb{Z}_2$. The Higgs state in the ${\mathbb P}_{1,1,2}$ fibration instead has charges $\left(+1,-2\right)$. Since it breaks a linear combination of the $U(1)$s we should change basis appropriately after which it can be seen that the breaking is now $U(1)_0\times U(1)_1 \rightarrow U(1)_2$ and there is no discrete gauge group remnant, cf.\ appendix~\ref{sec:discrete_subgroups_after_Higgsing}. 

The subtly which remains is how does the four-dimensional $\mathbb{Z}_2$ symmetry emerge form the three-dimensional $U(1)$ symmetry? To see this we should consider the states in the theory. Before the breaking we had a tower of KK states associated to the circle reduction. On the M-theory side the KK number maps to the wrapping number of the whole fiber of the M2-brane. In order to uplift a three-dimensional field to a four-dimensional one we require the full tower of KK states which span the full set of harmonic functions in the expansion of the wavefunction dependence of the four-dimensional field on the forth dimension. This translated to the fact that a KK $U(1)$ is a section, so each full fibre wrapping increased the KK number by just one. However after the Higgsing the remaining $U(1)_2=2U(1)_0+U(1)_1$, or in terms of the sections $S+Z$, is not longer a section and wrapping the fibre increases the KK charge by 2. If we look at the singly-charged states, those wrapping just the $A_1$-component or $B_1$-component of the fibre have odd charges. And the full KK, or wrapping, tower associated to them will therefore also have odd charges. If we use this tower to uplift a three-dimensional field to a four-dimensional one we can only recreate wavefunctions which are odd under reflection in the forth-direction. This acts as an effective $\mathbb{Z}_2$ symmetry. Integrating an odd number of such fields will yield zero.  

Note that the discussion above matches the torsion analysis of the manifolds. In the ${\mathbb P}_{1,1,2}$ fibration there is no torsion matching the absence of a discrete symmetry in three dimensions. While the corresponding Jacobian fibration has a torsion element matching the discrete gauge group in three dimensions.

\section{ \texorpdfstring{$SU(5) \times U(1)$}{SU(5)xU(1)} versus \texorpdfstring{$SU(5) \times \mathbb Z_2$}{SU(5)xZ2}}\label{sec:su5xu1-versus-su5Z2}

In this section we combine the $U(1)$ and $\mathbb Z_2$ symmetries studied previously with additional non-abelian gauge symmetry. 
Due to the appearance of additional charged matter and Yukawa interactions, we will be able to lend further support to the role of the $\mathbb Z_2$ symmetry as a discrete selection rule for the couplings of the theory.

Specifically, let us
implement an extra SU(5) gauge theory along a divisor $W: \theta =0$ on the base ${\cal B}$. Among the possible complex structure moduli restrictions giving rise to such a gauge group enhancement, a special class is given by \emph{toric tops} \cite{Candelas:1996su,Candelas:1997pv}. In this approach the base sections $b_i$ and $c_k$ factorise as $b_i = b_{i,j} \theta^j$ and $c_k = c_{k,l} \theta^l$ for suitable powers  $j$ and $l$ and $b_{i,j}$ and $c_{k,l}$ generic.
All such consistent configurations for all 16 torus fibrations realised as toric hypersurfaces have been classified in \cite{Bouchard:2003bu}, including their corresponding non-abelian gauge symmetries. Using the techniques of \cite{Bouchard:2003bu}, one finds that there are five such inequivalent specifications compatible with an $SU(5)$ gauge symmetry along $\theta=0$ for the $U(1)$ model \cite{Borchmann:2013hta,Borchmann:2013jwa,Braun:2013nqa} described in section \ref{sec:U(1)fibration}. For the $\mathbb Z_2$ model of section \ref{sec:Z2fibration} there are three inequivalent tops \cite{Braun:2013nqa}. But we should note here that the five tops in the $U(1)$ case can be matched by the three tops in the $\mathbb Z_2$ case by the additional symmetry of the fibre polygon after removing the point corresponding to the ambient fibre coordinate $s$.

\subsection{The \texorpdfstring{$SU(5) \times U(1)$}{SU(5)xU(1)} case}\label{sec:the-SU(5)xU(1)-case}

The details of the $SU(5) \times U(1)$ models have already been analysed in \cite{Borchmann:2013hta}, cf.~appendix A, but for the convenience of the reader we will repeat here the derivation of the most important results. 
We begin with the model described by the second $SU(5)$ top \cite{Borchmann:2013jwa} over polygon six in the enumeration by \cite{Bouchard:2003bu}, see Figure~\ref{fig:polygon6}. 
The proper transform of the hypersurface equation after resolution takes the form\footnote{Note the modified order of the exceptional divisors compared to \cite{Borchmann:2013hta}.} 
\begin{equation}\label{eq:hypersurface-su5xu1-model}
\begin{aligned}
P_1^{SU(5)}=&\,\,s w^2  e_1 e_2 + b_{0}  s^2 u^2 w e_0^2 e_1^2 e_2 e_4 + b_1s u v w  + b_2 v^2 w e_2 e_3^2 e_4\\
& + c_{0}s^3 u^4  e_0^4 e_1^3 e_2 e_4^2 + c_{1} s^2  u^3 v e_0^2 e_1 e_4 + c_{2}s u^2 v^2  e_0 e_3 e_4 + c_3 u v^3 e_0 e_2 e_3^3 e_4^2,
\end{aligned}
\end{equation}
where to avoid clutter we use $b_i$ and $c_k$ instead of $b_{i,j}$ and $c_{k,l}$ although we really mean the latter when referring to $b_i$ and $c_k$ in the sequel.

Using the hypersurface equation \eqref{eq:hypersurface-su5xu1-model} and the following Stanley-Reisner ideal 
\begin{equation}
\textmd{SR-i}:\quad\{vs, ve_1, ve_2, wu, we_0, we_4, ue_3, se_3, e_0e_3, e_1e_3, ue_1, 
ue_2, ue_4, se_4, e_1e_4, se_2\}\,,
\end{equation}
which corresponds to one of the phases of the resolution, we can work out the splitting of the fibre along the GUT divisor $W$. The five fibre components are given by
\begin{equation}
\begin{aligned}
\mathbb P^1_0 &= e_0 \cap b_1 s u v  +  s e_1 e_2 + b_2 v^2 e_2 e_4 \, , & w = e_3 = 1\,,\\
\mathbb P^1_1 &= e_1 \cap b_1 s w + b_2 w e_2 + c_{2} s e_0 + c_3 e_0 e_2 \, , & u = v =  e_3 = e_4 = 1\,,\\
\mathbb P^1_2 &= e_2 \cap b_1 w  + c_{1} e_0^2 e_1 e_4 + c_{2} e_0 e_3 e_4 \, , & u = v  = s=1\,,\\
\mathbb P^1_3 &= e_3 \cap b_{0} w e_2 e_4 +  b_1 v w + w^2 e_2 + c_{0} e_2 e_4^2 + c_{1} v e_4 \, , & u = s = e_0 = e_1  = 1\,,\\
\mathbb P^1_4 &= e_4 \cap b_1 v +  e_2  \, , & u = w = s= e_1 = 1\,.
\end{aligned}
\end{equation}

Due to the additional non-abelian singularity, the divisor \eqref{eq:Shiodamap} associated with the $U(1)$ symmetry of section \ref{sec:U(1)fibration} gets modified by the exceptional divisors $E_i$ of the $SU(5)$. The new $U(1)$ generator, which is uncharged under the non-abelian singularity, is given by 
\begin{equation}\label{eq:u1-generator-su5xu1-model}
\tw = 5(S-U-\bar{\mathcal{K}}-[b_2]) + 4 E_1 + 3 E_2 + 2 E_3 + E_4\,,
\end{equation}
where the overall normalisation has been chosen such as to render all appearing $U(1)$ charges integer in the sequel. 

\subsubsection{Matter curves}

To obtain the matter curves, we take the hypersurface equation \eqref{eq:hypersurface-su5xu1-model} prior to resolution\footnote{To obtain the original singular form of \eqref{eq:hypersurface-su5xu1-model} we just have to set $e_0$ to $\theta$ and $e_1$, $e_2$, $e_3$, $e_4$ to one.}  and calculate from it $f$ and $g$ of its associate Jacobian fibration. From $f$ and $g$ we can read off the discriminant $\Delta$ of the fibration. The divisor $\Delta=0$ gives the locus of the singular torus fibres. The vanishing order of $\Delta$ at that locus relates to the order of the singularity. We expand the discriminant in $\theta$,
\begin{equation}
\Delta \sim \theta^5 [b_1^4 b_2(b_1 c_3 - b_2 c_{2})(b_1^2 c_{0} - b_{0} b_1 c_{1} + c_{1}^2)  + \mathcal{O}(\theta)]\,,
\end{equation}
to look for singularity enhancements beyond $SU(5)$ along the GUT divisor. As we can see from the above equation, these are at
\[
\theta=b_1=0\,,\quad\theta= b_2=0\,,\quad\theta=(b_1 c_3 - b_2 c_{2}) =0\,,\quad\theta=(b_1^2 c_{0} - b_{0} b_1 c_{1} + c_{1}^2) =0\,.
\]
At these four curves matter transforming under the $SU(5)$ is localised. To determine the type of matter, one can either explore the vanishing orders of $f$ and $g$ at these curves or directly analyse the characteristics of the resolved fibres over these loci, which is the approach we will take in the following.

Along the curve\footnote{The the labeling of the curve will be justified a posteriori.} $\mathcal C_{\mathbf{10}_{-2}}=W \cap \{b_1=0\}$ the fibre components 
\begin{equation}
\begin{aligned}
\mathbb P^1_0 &= e_0 \cap  e_2(s e_1+ b_2 v^2 e_4),\\
\mathbb P^1_2 &= e_2 \cap  e_0 e_4(c_{1} e_0 e_1 + c_2 e_3 )
\end{aligned}
\end{equation}
factorise and the fibre topology becomes that of the affine $SO(10)$ Dynkin diagram. The intersection numbers of the new effective curves with the divisors $E_1$ up to $E_4$ are
\begin{equation} \label{10bar10}
\begin{aligned}
\mathbb P^1_{e_0 = e_2 = 0} \cdot (E_1, E_2, E_3, E_4) &= (1,-1,0,1)\,,\\
\mathbb P^1_{e_0 = se_1+b_2v^2e_4 = 0} \cdot (E_1, E_2, E_3, E_4) &= (0,1,0,0)\,.\\
\end{aligned}
\end{equation}
These intersection vectors are just the $U(1)$-Cartan charges of M2-branes wrapping these $\mathbb P^1$s. Therefore, they can be associated with states in the $\overline{\bf{10}}$ and ${\bf{10}}$ representation of $SU(5)$, respectively.

At the locus $\mathcal C_{\mathbf{5}_{-6}}=W \cap \{b_2=0\}$ the fibre curve
\begin{equation}
\mathbb P^1_0 = e_0 \cap  s(b_1 u v+ e_1 e_2)
\end{equation}
factorises. Calculating again the charges under the exceptional divisors, we find 
\begin{equation}
\begin{aligned}
\mathbb P^1_{e_0 = s = 0} \cdot (E_1, E_2, E_3, E_4) &= (1,0,0,0)\,,\\
\mathbb P^1_{e_0 = b_1 u v+ e_1 e_2 = 0} \cdot (E_1, E_2, E_3, E_4) &= (0,0,0,1)\,,
\end{aligned}
\end{equation}
which are the highest weights of the $\bf{5}$- and $\bar{\bf{5}}$-representation of $SU(5)$.

At the third enhancement locus, $\mathcal C_{\mathbf{5}_{4}}=W \cap \{b_1 c_3 - b_2 c_{2}=0\}$, we find the splitting
\begin{equation}
\mathbb P^1_1 = e_1 \cap  (c_3e_2+c_{2}s)(b_1 w+ c_{2} e_0)/c_{2}
\end{equation}
when solving for $b_2 = b_1 c_3 /c_{2}$ away from $\{c_{2}=0\}$. The charges under the exceptional divisors reveals again states in the $\bar{\bf{5}}$- and the $\bf{5}$-representation.

Finally, at the matter curve $\mathcal C_{\mathbf{5}_{-1}}=W \cap \{b_1^2 c_{0} - b_{0} b_1 c_{1} + c_{1}^2=0\}$ we find the splitting
\begin{equation}
\mathbb P^3_1 = e_3 \cap  (b_1 w+ c_{1} e_4)(b_1^2 v + b_{0} b_1 e_2 e_4 - c_{1} e_2 e_4)/b_{1}^2
\end{equation}
when solving for $c_{0}$ away from the locus $\{b_1=0\}$. These two new states correspond again to the fundamental and anti-fundamental representation of $SU(5)$.

The $U(1)$ charges for the matter states over the four curves $\mathcal C_{\mathbf{10}_{-2}}$, $\mathcal C_{\mathbf{5}_{-6}}$, $\mathcal C_{\mathbf{5}_{4}}$ and $\mathcal C_{\mathbf{5}_{-1}}$ are obtained by intersecting the new effective fibre components with the $U(1)$ generator $\tw$ given in \eqref{eq:u1-generator-su5xu1-model}. The intersections are $-2$, $-6$, $4$ and $-1$, respectively, thereby justifying the labeling of the curves.

Finally, we should note that the structure of $U(1)$ charged singlets is unaffected by the addition of the non-abelian gauge group factor along the divisor $W$, even though the specific form of the defining equations for the two types of singlet curves may differ slightly  compared to the pure $U(1)$ model. To derive the singlet curves we must take into account the appearance of factors of $\theta$ in $c_k = c_{k,l} \theta^l$ etc. The structure of the curves and their intersections is unchanged, though. Due to the overall---and arbitrary---normalization factor of $5$ in the $U(1)$ generator  \eqref{eq:u1-generator-su5xu1-model}, the singlets are now of charge $\mathbf{1}_{{10}}$ and $\mathbf{1}_{5}$.

\subsubsection{Yukawa couplings on \texorpdfstring{$W$}{W}}

There are three types of Yukawa points with couplings involving only the states charged under the $SU(5)$. The $\mathbf{10}_{-2}$-curve meets the $\mathbf{5}_{4}$- and the $\mathbf{5}_{-6}$-curves at $W \cap \{b_1=0\} \cap \{b_2=0\}$. The fibre enhances to the affine $SO(12)$ diagram at this locus. By grouping the irreducible fibre components one may construct a gauge singlet of states with the coupling $\mathbf{10}_{-2} \, \bar{\mathbf{5}}_6 \, \bar{\mathbf{5}}_{-4} + c.c.$

The $\mathbf{10}_{-2}$-curve intersects the $\mathbf{5}_{-1}$-curve at the points $W \cap \{b_1=0\} \cap \{c_{1}=0\}$. Over this locus the resolved fibre takes the form of the affine $SO(12)$ diagram. Constructing the gauge singlet identifies  the coupling $\bar{\mathbf{10}}_2 \, \mathbf{5}_{-1}\, \mathbf{5}_{-1} + c.c.$

The last Yukawa coupling between the $\mathbf{10}_{-2}$- and the $\mathbf{5}_4$-states are located at the point $W \cap \{b_1=0\} \cap \{c_{2}=0\}$. Here the enhancement type is $E_6$ and the invariant coupling is $\mathbf{10}_{-2} \, \mathbf{10}_{-2} \, \mathbf{5}_4 + c.c.$

In addition, three types of Yukawa couplings between the fundamental fields and the singlets occur:
At the intersection of the $\mathbf{1}_{\pm 10}$-curve with the GUT divisor $W$, {i.e.}~at $W \cap \{b_2=0\} \cap \{c_3=0\}$, the fibre enhances to an $SU(7)$. At this type of points the $\mathbf{5}_{-6}$- and $\mathbf{5}_4$-curve intersect, and by computing the charges of the split curves the Yukawa coupling $\mathbf{1}_{-10} \, \bar{\mathbf{5}}_6 \, \mathbf{5}_4 + c.c.$ is found.

At the intersection of the curves along which the $\mathbf{5}_4$ and $\mathbf{5}_{-1}$ are localised, which corresponds to  the points 
\begin{equation}
W \cap \{b_1 c_3 = b_2 c_{2}\} \cap \{b_1^2 c_{0} - b_{0} b_1 c_{1} + c_{1}^2 = 0\},
\end{equation}
the fibre enhances to $SU(7)$. Here the Yukawa coupling $\mathbf{1}_{-5}\, \mathbf{5}_4 \, \bar{\mathbf{5}}_1 + c.c.$ is localised.

At $W \cap \{b_2=0\} \cap \{b_1^2 c_{0} - b_{0} b_1 c_{1} + c_{1}^2=0\}$, where the $\mathbf{5}_{-6}$- and the $\bar{\mathbf{5}}_{1}$-curves intersect, the fibre looks again like an affine $SU(7)$ Dynkin diagram.  
Thus we have the coupling $\mathbf{1}_5\,\mathbf{5}_{-6}\,\bar{\mathbf{5}}_{1} + c.c.$. 

Finally, the universal ${\bf 1}_{10} {\bf 1}_{-5} {\bf 1}_{-5} + c.c.$ exists at the intersection of the two singlet curves, as in the model without $SU(5)$ enhancement.

\subsection{The \texorpdfstring{$SU(5) \times \mathbb Z_2$}{SU(5)xZ2} case}\label{sec:the-SU(5)xZ2-case}
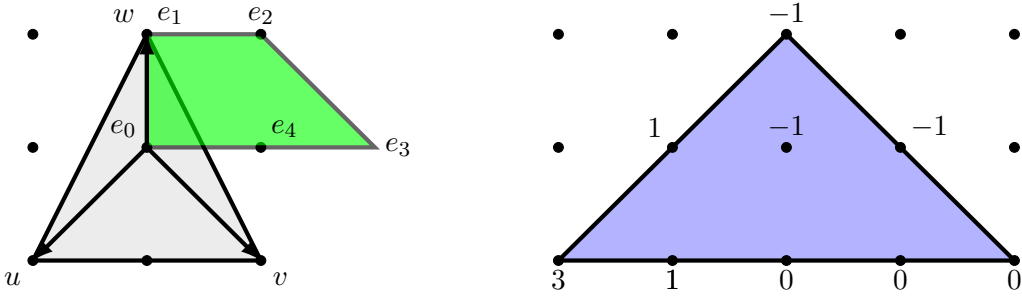
\begin{figure}[t]
    \centering
  \begin{tikzpicture}[scale=1.5]
  \filldraw [ultra thick, draw=black, fill=lightgray!30!white]
      (-1,-1)--(1,-1)--(0,1)--cycle;
    \foreach \x in {-1,0,1}{
      \foreach \y in {-1,0,...,1}{
        \node[draw,circle,inner sep=1.3pt,fill] at (\x,\y) {};
      }
    }
              \filldraw [ultra thick, draw=black, fill=green, opacity=0.6]
      (0,0)--(2,0)--(1,1)--(0,1)--cycle;
  \node [above left] at (0,0) {$e_0$};
    \node [above right] at (0,1) {$e_1$};
  \node [above] at (1,1) {$e_2$};
  \node [right] at (2,0) {$e_3$};
  \node [above right] at (1,0) {$e_4$};

  \draw[ultra thick, -latex]
       (0,0) -- (0,1) node[above left] {$w$};
  \draw[ultra thick, -latex]
       (0,0) -- (1,-1) node[below right] {$v$};
  \draw[ultra thick, -latex]
       (0,0) -- (-1,-1) node[below left] {$u$};
\begin{scope}[xshift=0.33\textwidth]
\filldraw [ultra thick, draw=black, fill=blue!30!white]
      (0,1)--(-2,-1)--(2,-1)--cycle;
    \foreach \x in {-2,-1,...,2}{
      \foreach \y in {-1,0,...,1}{
        \node[draw,circle,inner sep=1.3pt,fill] at (\x,\y) {};
      }
    }
  \node [above] at (0,0)  {$-1$};
  \node [above] at (0,1) {$-1$};
  \node [above right] at (1,0) {$-1$};
    \node [below] at (2,-1) {$0$};
      \node [below] at (1,-1) {$0$};
        \node [below] at (0,-1) {$0$};
          \node [below] at (-1,-1) {$1$};
   \node [below] at (-1,-1) {$1$};
   \node [below] at (-2,-1) {$3$};
  \node [above left] at(-1,0)  {$1$};
\end{scope}
  \end{tikzpicture}
      \caption{$SU(5)$ top over polygon 4 of \cite{Bouchard:2003bu} together with its dual polygon, bounded below by the values $z_{min}$, shown next to the nodes.}\label{fig:polygon4}
\end{figure}

In this subsection we will introduce an $SU(5)$ singularity for the $\mathbb Z_2$ model. Since we are interested in studying the relation of the $U(1)$ and the $\mathbb Z_2$ model via Higgsing, we will take the top which becomes the top of section \ref{sec:the-SU(5)xU(1)-case} after introducing the point corresponding to $s$. In the list of $SU(5)$ tops of \cite{Braun:2013nqa}, this is the third top over polygon four, denoted $\tau_{4,3}$. The proper transform of the hypersurface equation after resolving the $SU(5)$ singularity reads
\begin{equation}\label{eq:hypersurface-su5xZ2-model}
\begin{aligned}
P_2^{SU(5)}=&e_1 e_2 w^2 + b_0 u^2 w e_0^2 e_1^2 e_2 e_4 + b_1 u v w + b_2 v^2 w e_2 e_3^2 e_4\\
& + c_0 u^4 e_0^4 e_1^3 e_2 e_4^2 + c_1 u^3 v e_0^2 e_1 e_4 + c_2 u^2 v^2 e_0 e_3 e_4 + c_3 u v^3 e_0 e_2 e_3^3 e_4^2 + c_4 v^4 e_0 e_2^2 e_3^5 e_4^3\,,
\end{aligned}
\end{equation}
where we used again $b_i$ and $c_k$ instead of $b_{i,j}$ and $c_{k,l}$.

As in the $U(1)$ case, we work out the fibre components over the divisor $W$. They are given by
\begin{equation}
\begin{aligned}
\mathbb P^1_0 &= e_0 \cap b_1 u  + e_1 e_2 + b_2 e_2 e_3 \, , & v = w = e_3 = 1\,,\\
\mathbb P^1_1 &= e_1 \cap b_1 u w + b_2 w e_2 + c_2 u^2 e_0 + c_3 u e_0 e_2 + c_4 e_0 e_2^2 \, , & v =  e_3 = e_4 = 1\,,\\
\mathbb P^1_2 &= e_2 \cap b_1 w  + c_1 e_0^2 e_1 e_4 + c_2 e_0 e_3 e_4 \, , & u = v  = 1\,,\\
\mathbb P^1_3 &= e_3 \cap b_0 u^2 w e_2 e_4 +  b_1 u v w + w^2 e_2 + c_0 u^4 e_2 e_4^2 + c_1 u^3 v e_4 \, , & e_0 = e_1  = 1\,,\\
\mathbb P^1_4 &= e_4 \cap b_1 u +  e_2  \, , & v = w = e_1 = 1\,,
\end{aligned}
\end{equation}
where we used the SR-ideal
\begin{equation}
\textmd{SR-i}:\quad \{v\,e_0,\, v\,e_1,\, v\,e_2,\, w\,e_0,\, w\,e_4,\, u\,e_3,\, e_0\,e_3,\, e_1\,e_3,\, u\,e_2,\, e_1\,e_4,\, v\,w\,u\}\,
\end{equation}
corresponding to one of the phases of the resolution.
\begin{figure}
\vspace{-1cm}
\centering \def\svgwidth{350pt} 
\hspace{1cm}
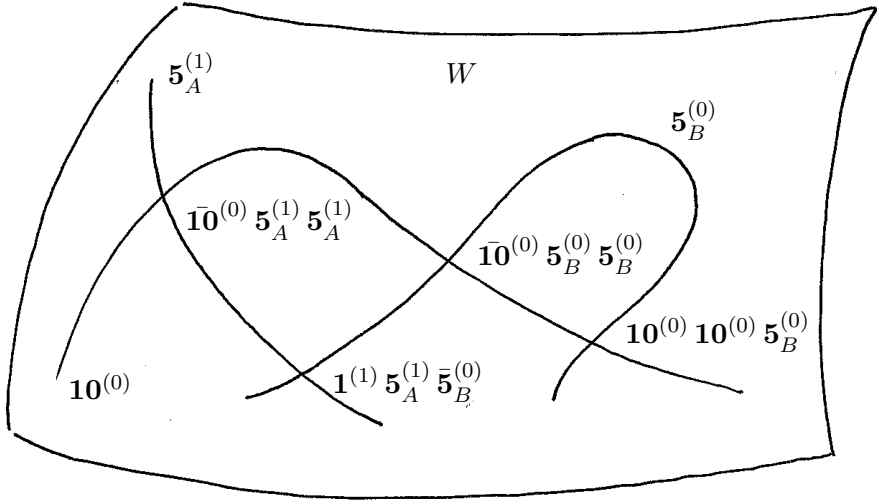 
\vspace{0cm}
\caption{The matter curves in $W: \{\theta = 0\}$ and the Yukawa couplings involving the $SU(5)$ charged matter in codimension three.}\label{fig:matter_curves_poly4}
 \end{figure}
\subsubsection{Matter curves}
Calculating again the discriminant of the associate Jacobian fibration to \eqref{eq:hypersurface-su5xZ2-model} and expanding it in $\theta$ yields
\begin{equation}
\Delta \sim \theta^5 [b_1^4(b_1^2 c_0 - b_0 b_1 c_1 + c_1^2)(b_2^2 c_2 - b_1 b_2 c_3 + b_1^2 c_4)  + \mathcal{O}(\theta)]\,.
\end{equation}
Interestingly, this time we only find three matter curves charged under the $SU(5)$. To identify the type of matter along these curves we redo the analysis of the last section.

At $\mathcal C_{\mathbf{10}}=W \cap \{b_1= 0\}$ the following fibre components split,
\begin{equation}
\begin{aligned}
\mathbb P^1_0 &= e_0 \cap  e_2(e_1+ b_2 e_3)\,,\\
\mathbb P^1_2 &= e_2 \cap  e_0 e_4(c_1 e_0 e_1 + c_2 e_3 )\,.
\end{aligned}
\end{equation}
For the intersection numbers with the exceptional divisors $E_i$ we find
\begin{equation}
\begin{aligned}
\mathbb P^1_{e_0 = e_2 = 0} \cdot (E_1, E_2, E_3, E_4) &= (1,-1,0,1)\,,\\
\mathbb P^1_{e_0 = e_1 + b_2 e_3= 0} \cdot (E_1, E_2, E_3, E_4) &= (0,1,0,0)\,,\\
\end{aligned}
\end{equation}
which are weight vectors in the $\bar{\mathbf{10}}$- and the $\mathbf{10}$-representation of $SU(5)$, respectively. The fibre topology is that of the affine $SO(10)$ diagram. For later purposes we also give the intersection with $U$. The bi-section intersects two of the $\mathbb P^1$'s with multiplicity one, specifically $\mathbb P^1_{1}|_{b_1 = 0}$ and $\mathbb P^1_{e_0 = e_1 + b_2 e_4 = 0}$.

The first fundamental matter curve, which we will call the $A$-curve in the following, is $\mathcal C_{\mathbf{5}_A}=W \cap \{b_1^2 c_0 - b_0 b_1 c_1 + c_1^2 = 0\} $, because along it $\mathbb P^1_3$ factorises as
\begin{equation}
\mathbb P^1_3 = e_3 \cap \frac{1}{b_1^2}(b_1 w + c_1 u^2 e_4)\big(b_1(b_1 u v + w e_2 + b_0 u^2e_2 e_4) -c_1 u^2 e_2 e_4) \big)\,.
\end{equation}
We used here that away from $\{b_1 = 0\}$ we may solve for $c_0$ and resubstitute back into the equations defining the fibre components. The intersection numbers for the two rational curves are
\begin{equation}
\begin{aligned}
\mathbb P^1_{e_3 = b_1 w + c_1 u^2 e_4 = 0} \cdot (E_1, E_2, E_3, E_4) &= (0,1,-1,0)\,,\\
\mathbb P^1_{e_3 = b_1(b_1 u v + w e_2 + b_0 u^2e_2 e_4) -c_1 u^2 e_2 e_4 = 0} \cdot (E_1, E_2, E_3, E_4) &= (0,0,-1,1)\,,\\
\end{aligned}
\end{equation}
which correspond to weight vectors associated with two states in the ${\bar{\mathbf{5}}}$- and the ${\mathbf{5}}$-representation, respectively. The fibre topology is that of the $SU(6)$ Dynkin diagram. The divisor class of the bi-section $U$ intersects the two adjacent nodes $\mathbb P^1_0$ and $\mathbb P^1_1$, each with multiplicity one. Note that these are roots.

Solving for $c_4$ along the third matter curve $\mathcal C_{\mathbf{5}_B}=W \cap \{b_2^2 c_2 - b_1 b_2 c_3 + b_1^2 c_4 = 0\} $ we find the factorisation
\begin{equation}
\mathbb P^1_1 = e_1 \cap \frac{1}{b_1^2}(b_1 u + b_2 e_2)\big(b_1(b_1 w + c_2 u e_0 + c_3 e_0 e_2) -b_2 c_2 e_0 e_2) \big)
\end{equation}
and the weights
\begin{equation}\label{eq:split_curves_at_5B}
\begin{aligned}
\mathbb P^1_{e_1 = b_1 u + b_2 e_2 = 0} \cdot (E_1, E_2, E_3, E_4) &= (-1,0,0,0)\,,\\
\mathbb P^1_{e_1 = b_1(b_1 w + c_2 u e_0 + c_3 e_0 e_2) -b_2 c_2 e_0 e_2 = 0} \cdot (E_1, E_2, E_3, E_4) &= (-1,1,0,0)\,.\\
\end{aligned}
\end{equation}
These again correspond to  states in the  ${\bar{\mathbf{5}}}$- and the ${\mathbf{5}}$-representation of $SU(5)$, respectively. The fibre forms again an $SU(6)$ structure over $\mathcal C_{\mathbf{5}_B}$. The divisor $U$ intersects the irreducible curves $\mathbb P^1_0$ and the second curve in \eqref{eq:split_curves_at_5B}. Thus, over the $B$-curve the bi-section intersects one of the two new effective curves responsible for the fundamental matter at this locus.

\subsubsection{The \texorpdfstring{$\mathbb Z_2$}{Z2}-charges of the states}

As explained already at the end of section \ref{sec:Z2fibration}, we can use the divisor $U$ to define a notion of $\mathbb Z_2$-charges for the singlets. As in the presence of a $U(1)$ gauge group we demand that the actual divisor whose intersection numbers with the fibre $\mathbb P^1$s wrapped by the associated M2-branes give the charges fulfils a suitable of horizontality condition, i.e. the intersection with the bi-section $U$ should vanish. Hence the appropriate divisor before adding the non-abelian singularities is not just $U$ but
\begin{eqnarray}
\tw_{\mathbb Z_2}=U-[b_2]+\bar {\cal K}\,.
\end{eqnarray}
Similarly to the divisors we usually obtain from sections via the Shioda map, we also demand that the intersections of such a divisor with all $\mathbb P^1$ fibres of the fibral divisors vanish, at least modulo two.  A divisor with this property is given by
\begin{equation}\label{eq:z2-generator-su5-model}
\tw_{\mathbb Z_2} + \tfrac45\,E_1+\tfrac35\,E_2+\tfrac25\,E_3+\tfrac15 E_4\,.
\end{equation}
For convenience we rescale the above divisor such as to achieve integer intersections with all rational lines and define 
\begin{equation}\label{eq:rescaled-z2-generator-su5-model}
Q_{\mathbb Z_2}:=5 \,\tw_{\mathbb Z_2} + 4\,E_1+3\,E_2+2\,E_3+E_4\,.
\end{equation}
The intersection numbers of this divisor with the five rational fibres of the divisors $E_0$ to $E_4$ are given by
\begin{equation}
Q_{\mathbb Z_2}\cdot (\mathbb P^1_0,\,\mathbb P^1_1,\,\mathbb P^1_2,\,\mathbb P^1_3,\,\mathbb P^1_4,\,)=(10,0,0,0,0)
\end{equation}
and thus vanish modulo $2 \times 5$, as demanded. Hence, $Q_{\mathbb Z_2}$ is a candidate to calculate a ${\mathbb Z_2}$-charge of the matter states.
To see this we calculate the intersection of $Q_{\mathbb Z_2}$  with the two fibral $\mathbb P^1$s over the \textbf{10}-curve defined in  (\ref{10bar10}) associated with the ${\bf \bar{10}}$ and ${\bf 10}$ representation,
\begin{equation}
Q_{\mathbb Z_2}\cdot ( \mathbb P^1_{e_0 = e_2 = 0}\,,\mathbb P^1_{e_0 = e_1 +b_2 e_3 = 0})= ( 2,\,8  ) = ( 2,\,-2 ), \quad\textmd{mod}\quad 10\,,
\end{equation}
its intersection with anti-fundamental and fundamental fibral $\mathbb P^1$s over the $\mathbf 5_A$-curve,
\begin{equation}
Q_{\mathbb Z_2}\cdot (\mathbb P^1_{e_3 = b_1 w + c_1 u^2 e_4 = 0} ,\, \mathbb P^1_{e_3 = b_1(b_1 u v + w e_2 + b_0 u^2e_2 e_4) -c_1 u^2 e_2 e_4 = 0})= ( 1,\,-1 ) \,,
\end{equation}
the corresponding intersections with the fibre over the $\mathbf 5_B$-curve,
\begin{equation}
Q_{\mathbb Z_2}\cdot ( \mathbb P^1_{e_1 = b_1 u + b_2 e_2 = 0},\, \mathbb P^1_{e_1 = b_1(b_1 w + c_2 u e_0 + c_3 e_0 e_2) -b_2 c_2 e_0 e_2 = 0}) = ( -4,\,4 ) \,,
\end{equation}
as well as the charges of the states over the singlet curve,
\begin{equation}
Q_{\mathbb Z_2}\cdot(\mathbb P^1_+,\,\mathbb P^1_-)=(5,\,5) =(-5,\,5)\quad\textmd{mod}\quad 10\,.
\end{equation}
As explained in appendix \ref{sec:discrete_subgroups_after_Higgsing}, these charges generate at first sight a $\mathbb Z_{10}$ symmetry, which however contains the center of $SU(5)$. To determine the actual discrete symmetry group realised in addition to the non-abelian $SU(5)$ we must correctly divide out this center. Following appendix \ref{sec:discrete_subgroups_after_Higgsing}, we can shift the discrete charges of the  fundamentals and antisymmetric states by $2\,n$ and $4\,n$ with $n\in \mathbb Z$, respectively, to find a canonical representative of $\mathbb Z_{10}/\mathbb Z_5$. Choosing $n=-2$ gives
\begin{equation}
\begin{aligned}
&(\bar{\mathbf{10}},\mathbf{10}):\, (10,\,-10)=(0,\,0)\quad\textmd{mod}\quad 10\,,\\
&(\bar{\mathbf 5}_A,\mathbf{5}_A):\,(5,-5)\quad \quad (\bar{\mathbf 5}_B,\mathbf{5}_B):\,(0,0).
\end{aligned}
\end{equation}
Recaling these charges by the inverse of the factor relating \eqref{eq:z2-generator-su5-model} and \eqref{eq:rescaled-z2-generator-su5-model} gives us the co-prime $\mathbb Z_2$-charges of the canonical representative of $\mathbb Z_{10}/\mathbb Z_5$.

Hence, $Q_{\mathbb Z_2}$ gives, as expected, well defined $\mathbb Z_2$-charges.
In the sequel we will denote the $\mathbb Z_2$ charges by a superscript (to distinguish them from the $U(1)$ charges prior to Higgsing).
The massless spectrum thus consists of the fields ${\bf 10}^{(0)}$, ${\bf 5}_A^{(1)}$, ${\bf 5}_B^{(0)}$ plus conjugates and the singlet ${\bf 1}^{(1)}$, see Figure~\ref{fig:matter_curves_poly4}.

\subsubsection{Yukawa points}

There is only one type of intersection points $W \cap \{b_1=0\} \cap \{c_1=0\}$  between the $\mathbf{10}^{(0)}$-curve and the fundamental $A$-curve. Here the fibre takes the form of an affine $SO(12)$ Dynkin diagram. From the fibre topology the coupling ${\mathbf{10}}^{(0)} \, {{\mathbf{\bar 5}}}_A^{(1)} \, {{\mathbf{\bar 5}}}_A^{(1)} + c.c.$ together is deduced. Clearly this is invariant under the assigned $\mathbb Z_2$ charges.

By contrast, the fundamental $B$-curve intersects the $\mathbf{10}^{(0)}$-curve at {\emph{two}} types of Yukawa points. At $W \cap \{b_1=0\} \cap \{b_2=0\}$ the fibre takes again the form of an affine $SO(12)$ Dynkin diagram. The Yukawa coupling here is the ${{\mathbf{10}}}^{(0)} \,{\bar{\mathbf{5}}}^{(0)}_B \, {\bar{\mathbf{5}}}^{(0)}_B + c.c.$.
At $W \cap \{b_1=0\} \cap \{c_2=0\}$ the fibre $\mathbb P^1$s intersect in the form of the non-affine $E_6$ Dynkin diagram. As we approach the points $W \cap \{b_1=0\} \cap \{c_2=0\}$ along the $\mathbf{10}^{(0)}$-curve, the following splitting occurs:
\begin{equation}
\begin{aligned}
\mathbb P^1_{e_2 = c_1 e_0 e_1 + c_2 e_3 = 0} &\rightarrow& \mathbb P^1_{e_2 = e_0 =0} \quad&+& \mathbb P^1_{e_2 = e_1 =0}\,,\\
(0,-1,0,1)\quad &\rightarrow& (1,-1,0,1)\quad &+& (-1,0,0,0)\,,\\
\mathbf{10}\qquad &\rightarrow& {\bar{\mathbf{10}}} \qquad&+& {\bar{\mathbf{5}}}_B\,.
\end{aligned}
\end{equation}
Following the logic of \cite{Marsano:2011hv} this gives a $\mathbf{10}^{(0)} \, \mathbf{10}^{(0)} \, {\mathbf{5}}^{(0)} _B + c.c.$ Yukawa coupling.

The intersection locus of the singlet locus $C$ with the fundamental matter curves can be shown to take the form $\mathcal C_{\mathbf{5}^{(1)}_A} \cap \mathcal C_{\mathbf{5}^{(0)}_B} \cap \{b_0b_2^2c_1c_2-b_1^2b_2c_0c_3 - b_2c_1^2c_3 + b_1^3c_0c_4 + b_1c_1^2c_4\}$ by using the prime ideal decomposition. It may be checked that this is a codimension three point lying in the GUT divisor. 
Consistently, the fibre over these points degenerates to form an $SU(7)$ Dynkin diagram. This indicates a Yukawa coupling ${\bf 5}^{(1)}_A \,{\bf \bar 5}^{(0)}_B \, {\bf 1}^{(1)}$.

\subsection{Interpretation}

The observed structure of matter curves and Yukawa interactions is indeed consistent not only with the appearance of a discrete $\mathbb Z_2$ selection rule for the $SU(5)$ model of section \ref{sec:the-SU(5)xZ2-case}, but in particular also with the interpretation of this selection rule precisely as the discrete remnant of the $U(1)$ gauge group realised in the $SU(5) \times U(1)$ fibration of section \ref{sec:the-SU(5)xU(1)-case} upon Higgsing along the $\mathbf{1}_{{10}}$ state. 
The $\mathbb Z_2$ selection rule manifests itself in the appearance of two distinct fundamental matter curves ${\cal C}_{{\bf 5}^{(1)}_A}$ and ${\cal C}_{{\bf 5}^{(0)}_B}$ and the fact that the corresponding states enjoy different couplings: After all, while the coupling $\mathbf{10}^{(0)} \, \mathbf{10}^{(0)}\, {\mathbf{5}}^{(0)}_B + c.c.$
 is realised, an analogous coupling of the form $\mathbf{10}^{(0)} \, \mathbf{10}^{(0)} \, {\mathbf{5}} ^{(1)}_A + c.c.$
 is absent from the geometry even though this coupling would be allowed on the basis of the $SU(5)$ symmetry. 
 This and the structure of the remaining Yukawas is consistent with our $\mathbb Z_2$ charge assignments.

Moreover, comparing the $SU(5) \times U(1)$ and the $SU(5) \times \mathbb Z_2$ models, the curve ${\cal C}_{{\bf 5}^{(0)}_B}$ is the result of recombining the matter curves ${\cal C}_{{\bf 5}_4}$ and ${\cal C}_{{\bf 5}_{-6}}$ upon Higgsing the $\mathbf{1}_{{10}}$ states, while the curve ${\cal C}_{{\bf 5}^{(1)}_A}$ and ${\cal C}_{{\bf 5}_3}$ are to be identified. Geometrically, if we un-Higgs the $\mathbb Z_2$ to $U(1)$ by setting $c_4=0$, the curve ${\cal C}_{{\bf 5}^{(0)}_B}$ splits into ${\cal C}_{{\bf 5}_4}$ and ${\cal C}_{{\bf 5}_{-6}}$. The recombination of the two curves upon Higgsing is possible due to the existence of the Yukawa coupling ${\bf  5}_{-6} {\bf \bar 5}_{-4} {\bf 1}_{10} + c.c.$. As ${\bf 1}_{10}$ develops a VEV, a holomorphic off-diagonal mass term for the fields ${\bf 5}_{-6} + c.c.$ and ${\bf 5}_{4}+ c.c.$ is induced such that only a single type of fundamental fields along the recombined locus remains.

Note that naively it might seem that due to the normalization of the $U(1)$ charges in presence of $SU(5)$ charged matter, the remnant discrete selection rule upon Higgsing the singlet field ${\bf 1}_{10}$ is $\mathbb Z_{10}$ and not $\mathbb Z_2$. However, a $\mathbb Z_5$ subgroup thereof is already accounted for by the center $\mathbb Z_5$ of the non-abelian $SU(5)$. In conclusion only an extra  $\mathbb Z_2$ selection rule remains in addition to the selection rules due to the $SU(5)$ gauge symmetry. The details of the embedding of the center group and how things can change if we go to other gauge groups in the $A$-series, we refer the reader to appendix~\ref{sec:discrete_subgroups_after_Higgsing}.

\section{R-parity by Higgsing a \texorpdfstring{$U(1)$}{U(1)} in F-theory}\label{sec:R-parity}

As another example we present, in this section, an $SU(5) \times \mathbb Z_2$ GUT model in which the discrete symmetry can be identified with R-parity.
The model is related via Higgsing to the $SU(5) \times U(1)$ model which has been constructed as 
 top 4 over polygon 6 in \cite{Borchmann:2013hta}. The hypersurface equation of this resolved $SU(5) \times U(1)$ fibration takes the form
\begin{equation}
\begin{aligned}
&w^2 s e_2 e_3^2 e_4  + b_0 s^2 u^2 w e_0 e_3 e_4 + b_1s u v w+ b_2 v^2 w e_1 e_2  \\
&+ c_0 s^3 u^4 e_0^3e_1e_3e_4^2 + c_1s^2 u^3ve_0^2e_1e_4 + c_2 s u^2v^2 e_0^2 e_1^2e_2 e_4 + c_3uv^3e_0^2e_1^3e_2^2e_4 = 0 \, .
\end{aligned}
\end{equation}
The Shioda map of the extra section is
\begin{equation}
W = 5(S-U-\bar{\mathcal{K}} - [b_2]) + \sum m_i E_i\,,\quad m_i = (2,4,6,3).
\end{equation}
Since the analysis of the charged matter representations and Yukawa couplings for this model has been performed in \cite{Borchmann:2013hta}, we merely restate the results here. Along $W \cap \{b_1=0\}$ the antisymmetric ${\bf 10}_{-1}$ is found. In addition, there are three fundamental matter curves. At $W \cap \{b_2=0\}$ the states in the ${\bf 5}_7 + c.c.$ are located. Along $W \cap \{b_1 c_0 - b_0 c_1=0\}$ the ${\bf 5}_2+c.c.$ states are found and over the curve $W \cap \{b_2^2 c_1 + b_1 c_2 - b_1^2 c_3=0\} $  ${\bf 5}_{-3} + c.c.$ matter is localised. 

There are two types of codimension-three enhancement points giving rise to Yukawa couplings among the $SU(5)$ charged matter. At $W \cap \{b_1=0\} \cap \{b_0=0\}$ the coupling ${\bf 10}_{-1}\,{\bf 10}_{-1}\,{\bf 5}_{2} + c.c.$ is located, and at $W \cap \{b_1=0\} \cap \{c_1=0\}$ the $\bar{\bf{10}}_{1}\,{\bf 5}_{-3}\,{\bf 5}_{2} + c.c.$ is found. There is also a non-flat point at $W \cap \{b_1=0\} \cap \{b_2=0\}$. The presence of this point has no effect on the following discussion. By an appropriate choice of the base for the fibration this point can be forbidden. 
In addition, all Yukawa couplings involving the singlets allowed by the $SU(5) \times U(1)$ gauge symmetry are indeed realised, specifically ${\bf 1}_{-10} \, {\bf 5}_7 \, {\bf \bar 5}_3$,      ${\bf 1}_{-5} \, {\bf  5}_7 \, {\bf \bar 5}_{-2}$,       ${\bf 1}_5 \, {\bf 5}_{-3} \,  {\bf \bar 5}_{-2}$ and of course ${\bf 1}_{10} \, {\bf 1}_{-5} \, {\bf 1}_{-5}$, plus their conjugates.



\begin{figure}[t]
    \centering
  \begin{tikzpicture}[scale=1.5]
  \filldraw [ultra thick, draw=black, fill=lightgray!30!white]
      (-1,-1)--(1,-1)--(0,1)--cycle;
    \foreach \x in {-1,0,1}{
      \foreach \y in {-1,0,...,2}{
        \node[draw,circle,inner sep=1.3pt,fill] at (\x,\y) {};
      }
    }
              \filldraw [ultra thick, draw=black, fill=green, opacity=0.6]
      (0,0)--(1,0)--(1,1)--(0,2)--cycle;
  \node [above left] at (0,0) {$e_0$};
    \node [above right] at (0,1) {$e_4$};
  \node [above right] at (1,1) {$e_2$};
  \node [left] at (0,2) {$e_3$};
  \node [above right] at (1,0) {$e_1$};

  \draw[ultra thick, -latex]
       (0,0) -- (0,1) node[above left] {$w$};
  \draw[ultra thick, -latex]
       (0,0) -- (1,-1) node[below right] {$v$};
  \draw[ultra thick, -latex]
       (0,0) -- (-1,-1) node[below left] {$u$};
\begin{scope}[xshift=0.33\textwidth]
\filldraw [ultra thick, draw=black, fill=blue!30!white]
      (0,1)--(-2,-1)--(2,-1)--cycle;
    \foreach \x in {-2,-1,...,2}{
      \foreach \y in {-1,0,...,1}{
        \node[draw,circle,inner sep=1.3pt,fill] at (\x,\y) {};
      }
    }
  \node [above] at (0,0)  {$-1$};
  \node [above] at (0,1) {$-1$};
  \node [above right] at (1,0) {$-1$};
    \node [below] at (2,-1) {$1$};
      \node [below] at (1,-1) {$1$};
        \node [below] at (0,-1) {$1$};
          \node [below] at (-1,-1) {$1$};
   \node [below] at (-2,-1) {$2$};
  \node [above left] at(-1,0)  {$0$};
\end{scope}
  \end{tikzpicture}
      \caption{$SU(5)$ top over polygon 4 of \cite{Bouchard:2003bu} together with its dual polygon, bounded below by the values $z_{min}$, shown next to the nodes.}\label{fig:Rparity_top}
\end{figure}
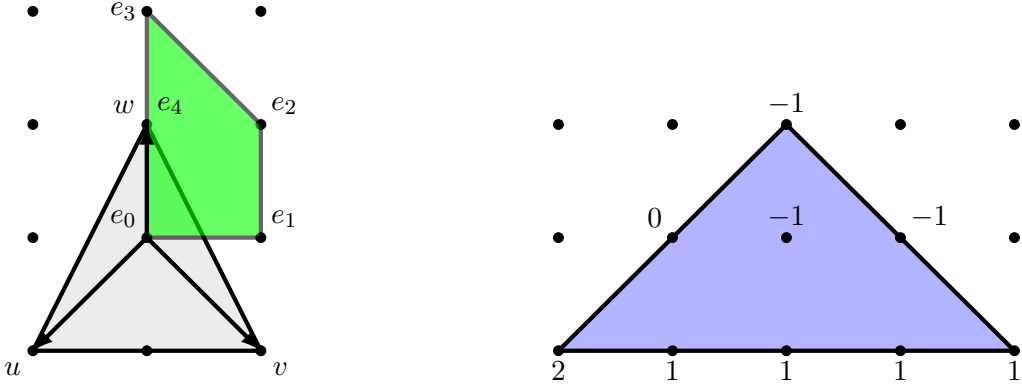

Giving a vacuum expectation value to the states in the ${\bf 1}_{\pm 10}$ representation breaks the $U(1)$ symmetry again  to a remnant $\mathbb Z_2$. 
The Higgsed model is described by the
 first top over polygon 4, denoted $\tau_{4,1}$ in \cite{Braun:2013nqa}, which gives for the hypersurface equation of the fourfold the following polynomial:
\begin{equation}
\begin{aligned}
&w^2 e_2 e_3^2 e_4 + b_0 u^2 w e_0 e_3 e_4 + b_1 u v w  + b_2 v^2 w e_1 e_2 \\
&+ c_0 u^4 e_0^3e_1e_3e_4^2 + c_1u^3ve_0^2e_1e_4 + c_2 u^2v^2 e_0^2 e_1^2e_2 e_4 + c_3uv^3e_0^2e_1^3e_2^2e_4 + c_4 v^4e_0^2e_1^4e_2^3e_4 =0.
\end{aligned}
\end{equation} 
From the class $U$ of the bi-section we construct the divisor
\begin{equation}
Q_{\mathbb Z_2} = 5\tw_{\mathbb Z_2} + 2E_1+4E_2+6E_3 + 3E_4, \qquad    \tw_{\mathbb Z_2}=U-[b_2]+\bar {\cal K}\,,
\end{equation} 
which has intersection number zero (modulo ten) with all fibre components in codimension one such that the roots are uncharged under the $\mathbb Z_2$ generator. 


Along the GUT divisor $W$ we now find only three, as opposed to four, matter curves. At $W \cap \{b_1=0\}$ the enhancement is of $SO(10)$ type, and by computing the intersection numbers of the split curves with the exceptional divisors $E_i$ we find weights of the anti-symmetric representation. Computing also the $\mathbb Z_2$ charges, which we denote again by a superscript, gives states in the ${\bf 10}^{(1)}$ and $\bar{{\bf 10}}^{(1)}$. 

Along the two curves at which the fundamental representations are localised the enhancement type is $SU(6)$. Along $W \cap \{b_1 c_0 - b_0 c_1=0\}$ we find states in the ${\bf 5}^{(0)} + c.c.$ These are the only invariant states under the action of $\mathbb Z_2$. Along the last matter curve $W \cap \{b_1 b_2^2c_2 - b_1^2b_2c_3 + b_1^3 c_4 - c_1b_2^3=0\}$ there is a ${\bf 5}^{(1)} + c.c.$ 
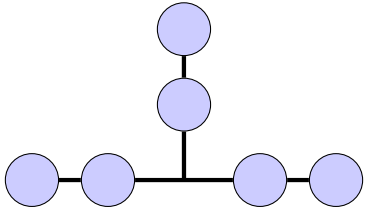
\begin{figure}[t]
    \centering
  \begin{tikzpicture}[scale=1.2,main node/.style={circle,fill=blue!20,draw,minimum size=7mm}]
  \node[main node] (1) {};
  \node[main node] (2) [left of=1] {};
  \node (3) [left of=2] {};
  \node[main node] (4) [left of=3] {};
  \node[main node] (5) [left of=4] {};
  \node[main node] (6) [above of=3] {};
  \node[main node] (7) [above of=6] {};
    \draw[ultra thick] (1) -- (2);
    \draw[ultra thick] (2) -- (3.center);
    \draw[ultra thick] (3.center) -- (4);
    \draw[ultra thick] (4) -- (5);
    \draw[ultra thick] (3.center) -- (6);
    \draw[ultra thick] (6) -- (7);
  \end{tikzpicture}
      \caption{Fibre topology at the Yukawa point $W\cap \{b_0=0\} \cap \{b_1=0\}$.}\label{fig:Rparity_yukawa_nonstandard}
\end{figure}

There are altogether three types of enhancement points in codimension three. 
At $W \cap \{b_0=0\} \cap \{b_1=0\}$ the ${\bf 10}_1$- and the ${\bf 5}_0$- curve intersect. Here the fibre topology is of non-standard, $E_6$-like, form where the three inner nodes all intersect in one point, see Figure~\ref{fig:Rparity_yukawa_nonstandard}. At this point the coupling ${\bf 10}^{(1)}\, {\bf 10}^{(1)}\, {\bf 5}^{(0)} + c.c.$ is localized. 
The second Yukawa coupling is found at $W\cap \{b_1=0\} \cap \{c_1=0\}$, which is a point of $SO(12)$ enhancement. This is where all the  three $SU(5)$-charged matter curves meet, and we confirm the coupling $\bar{{\bf 10}}^{(1)} \, {\bf 5}^{(0)} \, {\bf 5}^{(1)} + c.c.$ from the fibre topology. 
Finally, the two distinct ${\bf 5}$-curves intersect at the point ${\cal C}_{{\bf 5}^{(0)}}\cap {\cal C}_{{\bf 5}^{(1)}} \cap \{b_2^3c_0 - b_0b_2^2c_2 + b_0b_1b_2c_3 - b_0b_1^2c_4=0\}$. This is computed as the prime ideal decomposition of the intersection of the singlet locus at $W$ with the fundamental matter curves. Here the Yukawa coupling ${\bf 1}^{(1)}  \, {\bf 5}^{(0)} \, {\bf \bar 5}^{(1)} + c.c. $ arises.
Note that at the point $W \cap \{b_1=0\} \cap \{b_2=0\}$ the fibre is again non-flat, and this point must be absent in order for the fibration to give rise to a well-defined F-theory compactification. This can be achieved by choosing a base space $\mathcal B$ with specific intersection properties.

The interpretation of this spectrum and the interactions is again consistent with the origin of the $\mathbb Z_2$ as a discrete subgroup of the Higgsed $U(1)$. The Higgsing recombines the curves with states ${\bf 5}_7$ and ${\bf 5}_{-3}$, which couple to the Higgs field via the Yukawa  ${\bf 1}_{-10} \, {\bf 5}_7 \, {\bf \bar 5}_3$, into a single curve with states of $\mathbb Z_2$-charge $1$. This is evident by noting that this latter curve factorises accordingly as we un-Higgs the $U(1)$ by setting $c_4 =0$. All other curves are unaffected (since they do not couple to the Higgs field) and the $\mathbb Z_2$ charges of the states after the transition equal the former $U(1)$ charges mod 2.
The realised Yukawa couplings respect this $\mathbb Z_2$ symmetry and are related to the Yukawa couplings in the $SU(5) \times U(1)$-model as expected upon Higgsing.

Interestingly, the $\mathbb Z_2$ selection rule realised in this $SU(5)$ GUT model coincides precisely with matter $R$-parity:
The only field with trivial $\mathbb Z_2$ charge is the ${\bf 5}^{(0)}+c.c.$, which is consequently identified with ${\bf 5}_{H^u} + {\bf \bar 5}_{H^d}$ field. The non-trivial representations under $\mathbb Z_2$ are taken as the GUT matter representations, in particular the ${\bf \bar 5}^{(1)}$ is identified with ${ \bf \bar 5}_m$ and the singlet ${\bf 1}^{(1)}$ corresponds to the right-handed neutrino. The singlet coupling ${\bf 1}^{(1)}  \, {\bf 5}^{(0)} \, {\bf \bar 5}^{(1)}$ thus describes a Dirac mass for the right-handed neutrinos.

\section{Conclusions}\label{sec:conclusions}

In this paper we have studied the realisation of discrete gauge symmetries in F-theory compactifications to four dimensions via Higgsing. 
In the setup we have considered, a discrete $\mathbb{Z}_2$ symmetry originates as the remnant of a $U(1)$ gauge symmetry  upon Higgsing the latter by a field of charge 2. This amounts to a deformation \cite{Morrison:2014era} of the generic elliptic fibration with two sections \cite{Morrison:2012ei} (and thus a $U(1)$ gauge symmetry) into a bi-section fibration \cite{Braun:2014oya}. We have studied this process in detail focusing on aspects which are new to four-dimensional compactifications and in the presence of an additional non-abelian gauge group. We have shown that the Higgsing induces matter curve recombination, and that the resulting curves can be associated a $\mathbb{Z}_2$ charge through a generalisation of the Shioda map for multi-sections. We have further shown that the induced $\mathbb{Z}_2$ charge implies a selection rule on Yukawa couplings which leads to couplings being absent in the geometry even without a $U(1)$ symmetry forbidding them.  This is the first implementation of these aspects of discrete symmetries in a semi-realistic F-theory compactification. In particular we have presented an $SU(5)$ model with $\mathbb{Z}_2$ charge assignments which are equivalent to R-parity in the MSSM. It would be very interesting to apply this technology to induce other phenomenologically desirable discrete symmetries.

The Higgsing process has also been shown to induce an associated $G_4$-flux which compensates for the change in the Euler number of the fourfold such that the D3 tadpole is left invariant.  
The flux we have found is very similar to the one that has described before in conifold transitions of the $U(1)$ restricted model \cite{Grimm:2010ez}, which is a special case of the present geometries.
The conifold transition for this model had been studied in \cite{Braun:2011zm} and the fluxes found therein have been generalized to models with extra non-abelian singularities in \cite{Krause:2012yh}.
It will be interesting to generalize the fluxes found in this paper for the conifold transition at hand to the models with extra non-abelian singularities, e.g.\ of type $SU(5)$, along the lines of \cite{Krause:2012yh}.

Finally we have shown an explicit map between a three-dimensional field theory description of the Higgsing and the geometry. This involves an interesting interplay between the Coulomb branch and Kaluza-Klein modes. Further, we have been able to calculate the mixing of the three-dimensional Kaluza-Klein gauge field with the $U(1)$ field by calculating the charges of the Higgsed state.

\subsubsection*{Acknowledgements}

\noindent We thank Arthur Hebecker and Ling Lin for important discussions. The work of EP is supported by the Heidelberg Graduate School for Fundamental Physics. Furthermore, this work was  supported in part by the DFG under TR 33 `The Dark Universe'.

\appendix
\section{Discrete subgroups after Higgsing}\label{sec:discrete_subgroups_after_Higgsing} \label{App-disc}
In this appendix we derive in detail the remnant discrete subgroup after Higgsing a $U(1)$ in the presence of matter charged under another `spectator' gauge group. 
We exemplify our general results for additional $U(1)$
or $SU(N)$ spectator groups as appearing in the recent F-theory literature. 

As is well-known, if we give a VEV to a field transforming only under a $U(1)$ with charge $q^H$, we Higgs the $U(1)$ gauge symmetry to $\mathbb Z_{q^H}$. However, this is only true if the $U(1)$-charges of all the fields $\{\varphi_I\}$\footnote{Note that this set also includes the Higgs $H$, i.e.\ the field which obtains the VEV. In the sequel we will sometimes treat the Higgs field, for presentational purposes, separately. In this case we will use a lower case $i$ to denote all fields different from the Higgs.}  charged under the $U(1)$ are properly normalised or co-prime\footnote{Note that in the full theory there are also line operators which define a quantised unit charge independent of matter fields. However in the following analysis we are concerned with discrete symmetries acting on matter fields and with such symmetries there is not a notion of an absolute charge but only a relative one. } , i.e.\ $\textmd{GCD}(\{q^{I}\})=\textmd{GCD}(\{q^H,q^{i}\})=1$. Therefore, the actual discrete symmetry is $\mathbb Z_{q^H/\textmd{GCD}(\{q^{I}\})}$. To prevent cumbersome notation we will use capital letters for co-prime charges, i.e.\ $Q^{I}=q^{I}/\textmd{GCD}(\{q^{I}\})$.

Further subtleties can arise if some of the fields $\{\varphi_i\}$ transform in non-trivial representations of other abelian or non-abelian gauge symmetries $G_r$ (with abelian discrete subgroups). This is because a subgroup $\mathbb Z_{N_\textmd{sub}}$ of  $\mathbb Z_{Q^H}$ might be part (or all) of the discrete abelian subgroups $\mathbb Z_{N_r}$ of $G_r$ and thus needs to be divided out to avoid double-counting. In such a situation the actual remaining symmetry group after Higgsing is
\[
G_r\times  \frac{\mathbb Z_{Q^H}}{\mathbb Z_{N_\textmd{sub}}}.
\]
To obtain the subgroup $\mathbb Z_{N_\textmd{sub}}$ we first consider the subgroup $\mathbb Z_{N_\textmd{tsg}}$ of $\mathbb Z_{Q^H}$ which acts trivially on all the fields $\{\varphi_\alpha\}\subset \{\varphi_I\}$ which are uncharged under $G_r$.  The generator of $\mathbb Z_{N_\textmd{tsg}}$ is given by taking
\[
\textmd{LCM}\left(\left\{\frac{\textmd{LCM}(Q^\alpha,Q^H)}{Q^\alpha}\right\}\right)
\]
times the generator of $\mathbb Z_{Q^H}$, i.e.
\[
N_\textmd{tsg}=\frac{Q^H}{\textmd{LCM}\left(\left\{\frac{\textmd{LCM}(Q^\alpha,Q^H)}{Q^\alpha}\right\}\right)}.
\]
 If $\textmd{GCD}(N_\textmd{tsg},N_r)\neq 1$ then there can be a subgroup $\mathbb Z_{N_\textmd{sub}}$ of both $\mathbb Z_{N_\textmd{tsg}}$ and $\mathbb Z_{N_r}$ such that the elements of the representations of $\mathbb Z_{N_\textmd{tsg}}$ and $\mathbb Z_{N_r}$ agree on this subgroup $\mathbb Z_{N_\textmd{sub}}$.
In this case $N_{\rm sub}$ is the order of this subgroup. If $\textmd{GCD}(N_\textmd{tsg},N_r)= 1$, there cannot be a common non-trivial subgroup of both $\mathbb Z_{N_\textmd{tsg}}$ and $\mathbb Z_{N_r}$.

To be more specific about the identification of $\mathbb Z_{N_\textmd{sub}}$, we will now consider two explicit examples, given by $G_r =SU(N)$ with $N=4,5$ and, respectively, $G_r =U(1)$. 
For the first example with $G_r=SU(N)$, we specify the matter content,  i.e.\ the set of fields $\{\varphi_I\}$, as appearing in the theories realised (by top constructions) in \cite{Borchmann:2013hta,Borchmann:2013jwa},
\begin{equation}
\mathbf 1_{2\,N},\qquad \mathbf 1_N,\qquad {\yng(1,1)}_{\,2\,n},\qquad \left\{{\yng(1)}_{\,n+i\,N}\right\}_{i\in S_i}\
\end{equation} 
with $n=0,\ldots,N-1$ for the different tops, $S_i$ some `integer interval' which has zero as an element, cf.~\cite{Braun:2013nqa}, and {\tiny \yng(1)} and {\tiny \yng(1,1)} the fundamental and anti-symmetric representation, respectively. The subscripts next to the states denote the charge under the $U(1)$ which is Higgsed by giving $\mathbf 1_{2\,N}$ a VEV. $Q^H$ is therefore
\begin{equation}
Q^H = \left\{
  \begin{array}{rl}
    10 &:  N=5,\,n=1,\ldots,4\\
    2 & :  N=5,\,n= 0
  \end{array}
\right.\,,\qquad
Q^H = \left\{
  \begin{array}{rl}
    8 & : N=4,\,n=1,3\\
    4 & : N=4,\,n=2\\
    2 & : N=4,\,n=0
  \end{array}
\right.\,.
\end{equation}
Since there is only one additional $SU(N)$ singlet with half the $U(1)$-charge of the Higgs, one concludes $N_\textmd{tsg} = \frac{1}{2} Q^H$. Hence for $n=0$, $N_{\rm sub} =1$. For the rest we have to work a bit more.
The action of $\mathbb Z_{N_\textmd{tsg}}$ on the fields can be identified by with a $\mathbb Z_{\frac{1}{2} Q^H}$ action with generator
\begin{equation}
e^{2\pi i\frac{2 Q^i}{Q^H}} \quad\textmd{with}\quad \left(2\tfrac{Q^{\tiny{\yng(1,1)}_{2\,n}}}{Q^H},2\tfrac{Q^{{{\tiny\yng(1)}}_{n}}}{Q^H} \right)\cong \left\{
  \begin{array}{rl}
  (\frac{2n}{N},\frac{n}{N})=(\frac{2n}{5},\frac{n}{5}) & : N=5,\,n=1,\ldots,4\\
  (\frac{2n}{N},\frac{n}{N})=(\frac{2n}{4},\frac{n}{4}) & : N=4,\,n=1,3\\
  (\frac{2n}{N},\frac{n}{N})\cong (0,\frac12)\,\,\,\, & : N=4,\,n=2
  \end{array}
\right.\,.
\end{equation}
We have given only one generator of $\mathbb Z_{N_\textmd{tsg}}$ for the fundamentals because they are the same for all $i$'s. 
In the first two cases $\mathbb Z_{N_\textmd{tsg}}$ agrees with the center of $SU(5)$ and $SU(4)$, respectively. In the last case only the $\mathbb Z_2$ subgroup of the $SU(4)$ center is generated.
Hence we obtain 
\begin{equation}
 \frac{\mathbb Z_{Q^H}}{\mathbb Z_{N_\textmd{sub}}} = \left\{
  \begin{array}{rl}
    \mathbb Z_{10}/\mathbb Z_5 &:  N=5,\,n=1,\ldots,4\\
    \mathbb Z_2 & :  N=5,\,n= 0
  \end{array}
\right.\,,\qquad
 \frac{\mathbb Z_{Q^H}}{\mathbb Z_{N_\textmd{sub}}} = \left\{
  \begin{array}{rl}
    \mathbb Z_8/\mathbb Z_4 & : N=4,\,n=1,3\\
    \mathbb Z_4/\mathbb Z_2 & : N=4,\,n=2\\
    \mathbb Z_2 & : N=4,\,n=0
  \end{array}
\right.\,.
\end{equation}
We find that the discrete part is for all examples $\mathbb Z_2$. Its realisation depends however strongly on the matter content. For $\mathbb Z_{10}/\mathbb Z_5$ and $\mathbb Z_8/\mathbb Z_4$ there is a (canonical) representative within the equivalence class generating the $\mathbb Z_2$, which acts  either with $1$ or $-1$ on all the matter states---like for the $n=0$ cases where the charges are right from the beginning either zero or one-half. However there is no such representative for $N=4$ and $n=1,3$.\footnote{Repeating the same analysis for a spectator $SU(2)$ with charge one modulo two for the fundamentals, one finds a remaining $\mathbb Z_4/\mathbb Z_2$-symmetry upon switching on a VEV for the singlet of charge four.  This situation describes a transition from the $\textmd{Bl}^1\mathbb P_{[1,1,2]}$-fibration to a $\mathbb P_{[1,1,2]}$-fibration in the presence of an $SU(2)$. Hence one would naively expect a $\mathbb Z_4$ after Higgsing, cf.~\cite{Klevers:2014bqa}, but as the above analysis shows the actual discrete abelian group is just  $\mathbb Z_2$.}

Our second example is a  $U(1)_a\times U(1)_b$-model with the matter content
\begin{equation}\label{eq:states-U1xU1-model}
\mathbf 1_{0,2},\qquad \mathbf 1_{-1,-2},\qquad \mathbf 1_{1,-1},\qquad \mathbf 1_{1,0},\qquad \mathbf 1_{-1,-1},\qquad \mathbf 1_{0,1}\,.
\end{equation} 
This theory is realised by F-theory on the elliptic fibrations studied in \cite{Borchmann:2013hta,Cvetic:2013nia,Cvetic:2013uta,Borchmann:2013jwa}.
We Higgs this model in two different ways. In the first case we will give $\mathbf 1_{0,2}$ a VEV and in the second case we switch on a VEV for $\mathbf 1_{-1,-2}$. For $\left\langle\mathbf 1_{0,2}\right\rangle\neq 0$ the situation is pretty obvious. First of all we note that all the charges are co-prime. Secondly, the Higgs $\mathbf 1_{0,2}$ is only charged under $U(1)_b$. Hence, $G_r=U(1)_a$ and $\mathbb Z_{Q^H}=\mathbb Z_2$. There is one other field $\mathbf 1_{0,1}$ which is only charged under the second factor. Hence,  $N_\textmd{tsg}=1$ and
\[U(1)_a\times\mathbb Z_2\]
is the remaining symmetry. This agrees with the model considered in ~\cite{Klevers:2014bqa}.

Alternatively let us consider a Higgsing with $\left\langle\mathbf 1_{-1,-2}\right\rangle\neq 0$. In this case, we have to choose a different basis for the $U(1)$s. For general $q_a^H$ and $q_b^H$, the direction which leaves $\varphi_H$ invariant is 
\begin{equation}\label{eq:u1-direction-one}
\phi_a=\frac{\textmd{LCM}(q_a^H,q_b^H)}{q_a^H}\phi_{a'}\,,\qquad\phi_b=-\frac{\textmd{LCM}(q_a^H,q_b^H)}{q_b^H}\phi_{a'}\,.
\end{equation}
The second direction  we choose such that it generates together with \eqref{eq:u1-direction-one} the $\mathbb Z^2$ charge-lattice of $U(1)_a\times U(1)_b$, i.e.
\begin{equation}\label{eq:u1-direction-two}
\phi_a=D\,\phi_{b'}\,,\quad \phi_b=C\,\phi_{b'}\quad\textmd{with}\quad C\,\frac{\textmd{LCM}(q_a^H,q_b^H)}{q_a^H}+D\,\frac{\textmd{LCM}(q_a^H,q_b^H)}{q_b^H}=1\,.
\end{equation}
Hence, we obtain $(2,-1)$ for the $U(1)_{a'}$ direction and for $U(1)_{b'}$ we choose $(-1,0)$ out of the possible solutions to \eqref{eq:u1-direction-two}.
The states \eqref{eq:states-U1xU1-model} read as follows in the new $U(1)_{a'}\times U(1)_{b'}$ basis:
\begin{equation}
\mathbf 1_{-2,0},\qquad \mathbf 1_{0,1},\qquad \mathbf 1_{3,-1},\qquad \mathbf 1_{2,-1},\qquad \mathbf 1_{-1,1},\qquad \mathbf 1_{-1,0}\,.
\end{equation}
Since all the $Q^I$ charges are co-prime and $Q^H=1$,  the remaining symmetry is just $U(1)_{a'}$.


\bibliography{papers}  
\bibliographystyle{custom1}

\end{document}